\documentclass{elsart}

\usepackage{epsfig}

\usepackage{amssymb,graphicx}

\begin{document} 

\hyphenation{in-ho-mo-ge-neous}

\begin{frontmatter}



\title{Interface Conditions for Wave Propagation Through Mesh Refinement Boundaries}


\author[LHEA,USRA]{Dae-Il Choi},
\ead{choi@milkyway.gsfc.nasa.gov}
\author[LHEA,NCSU]{J. David Brown\thanksref{NRC}},
\ead{david\_brown@ncsu.edu}
\author[LHEA,UMD]{Breno Imbiriba},
\ead{imbiriba@milkyway.gsfc.nasa.gov}
\author[LHEA]{Joan Centrella},
\ead{jcentrel@milkyway.gsfc.nasa.gov}
\author[Drexel]{Peter MacNeice}
\ead{macneice@alfven.gsfc.nasa.gov}

\thanks[NRC]{Senior NRC Associate}

\address[LHEA]{Laboratory for High Energy Astrophysics, NASA/Goddard Space Flight Center,
Greenbelt, MD 20771 USA}
\address[USRA]{Universities Space Research Association, 7501 Forbes Boulevard, \#206, Seabrook,
MD 20706 USA}
\address[NCSU]{Department of Physics, North Carolina State University, Raleigh, NC 27695 USA}
\address[UMD]{Department of Physics, University of Maryland, College Park, MD 20742 USA}
\address[Drexel]{Department of Physics, Drexel University, Philadelphia, PA 19104 USA}

\begin{abstract}
We study the propagation of waves across fixed mesh refinement boundaries in 
linear and nonlinear model equations in 1--D and 2--D, and in the 3--D Einstein 
equations of general relativity. We demonstrate that using linear interpolation 
to set the data in guard cells leads to the production of reflected waves at 
the refinement boundaries. Implementing quadratic interpolation to fill the 
guard cells suppresses these spurious signals. 
\end{abstract}

\end{frontmatter}

\section{Introduction}
\label{intro}

Wave propagation is an important phenomenon throughout all areas of physics, 
with applications
typically involving multiple spatial and temporal scales. In numerical modeling 
of such problems, one strategy for dealing with the disparity in spatial and 
temporal scales is the use of a nonuniform or adaptive computational mesh. 
In this case waves can cross mesh refinement boundaries as they propagate 
through the computational domain.  This paper focuses on interface
conditions that will allow waves to travel smoothly across fixed refinement 
boundaries, minimizing spurious reflections.

The specific application that motivated this study is modeling the emission
of gravitational waves from astrophysical sources such as binary black hole
and neutron star coalescences. 
Such systems are among the most important sources for ground-based gravitational
wave detectors such as LIGO and VIRGO \cite{LIGO,LIGO2}, as well as the planned
space-based LISA mission \cite{LISA}. The gravitational waves produced
typically have wavelengths $\sim 10 - 100$ times the scales of their sources
Numerical simulations of such systems must therefore allow the signals to
propagate from finely resolved regions around the sources into more coarsely
resolved regions in the wave zones. Since the waveforms must be computed
at large distances from their sources ({\em i.e.}, effectivelly  at infinity)
for comparison with observations from gravitational wave detectors,
the simulation domains must be made as large as possible. 
This can be achieved by incorporating several levels of successively
coarser grids.

The propagation of gravitational waves is governed by the Einstein equations,
which are a coupled set of nonlinear partial differential equations \cite{MTW}. 
These equations can be written in a variety of ways, but current practice
in numerical relativity favors the use of the so--called BSSN formalism \cite{SN,BS}.
In this formalism, the Einstein equations are written as a system of quasi-linear
equations with first--order time derivatives and second--order spatial derivatives.
In this paper we restrict our analysis to the ``iterated Crank--Nicholson''
update scheme, which is a second order accurate, explicit finite difference method
that is currently in widespread use in the relativity community. It should
also be noted that we consider mesh refinement only in space, not in time.
In particular, for our present analysis we use a common timestep across 
the entire computational domain. 

Adaptive mesh refinement (AMR) was first applied in numerical relativity
to study critical phenomena in the 1--D collapse of a scalar field
to form a black hole \cite{choptuik93}. An early 3--D application focussed
on evolving a single black hole \cite{slicebh}; this was followed
by the use of fixed mesh refinement (FMR) to evolve a short part of
a binary black hole evolution \cite{brugbbh}. AMR was also employed
to follow the propagation of gravitational waves through spacetime,
first using a single model equation that describes perturbations of
a non-rotating black hole \cite{PSW} and later in the 3--D Einstein 
equations \cite{Newetal}, and to study inhomogeneous cosmological
models \cite{hern-thesis}. In these AMR studies the refinement and
derefinement conditions were generally tuned so that the gravitational
waves remained within the finely resolved regions.

In this paper, we address the challenge of evolving gravitational
wave signals {\em across} mesh refinement boundaries using FMR.
Success in this endeavor is an essential component of gravitational
wave source modeling, due to the large disparity in the scales of
the sources and the waves.
Our challenge amounts to choosing a prescription for coupling adjacent
grid blocks when the blocks have different resolutions.
Grid blocks are coupled through their guard cells, which must be filled
using data from the blocks' interior cells. In hydrodynamics codes
it is common practice to use a linear interpolation scheme for guard
cell filling, with a possible adjustment for flux conservation across
the interface between blocks \cite{BergerOliger,BergerColella,berger87}.
We have found that this prescription is not adequate for the BSSN
formulation of the Einstein equations. In particular, linear guard cell
filling leads to unacceptably large reflections and distortions of
the gravitational waves as they propagate from fine grid blocks
to coarse grid blocks.
Our solution to this problem is to use a guard cell filling procedure
with quadratic--order accuracy orthogonal to the coarse--fine grid
interface.
The need for quadratic order guard cell filling has been previously
demonstrated for elliptic boundary value problems with second order
derivatives in \cite{Chesshire90,Henshaw03}.
With this prescription spurious wave reflections and distortions
are reduced dramatically.

Given the complexity of the full system of Einstein equations,
we have chosen to analyze first a set of model wave equations
in 1-D and 2-D that mimic some of the properties of the Einstein equations,
as expressed in BSSN form. These simplified test beds have proved essential 
to understanding and correcting the problems that arise in the propagation
of waves across mesh refinement boundaries. Since the solution we uncovered
using these model equations has proved effective in curing the difficulties
encountered in the Einstein equations, we expect this work to be useful
across a broad range of related wave propagation problems.

\section{Linear Wave Equation in 1-D: Evolution on a Uniform Grid}
\label{lin-1d}

The linear wave equation in 1-D is generally written in the form
\begin{equation}
{\partial^2 \phi \over \partial t^2} =  {\partial^2 \phi \over \partial x^2},
\label{1-d-std}
\end{equation}
where $\phi = \phi(x,t)$.  Introducing the auxiliary variable $\Pi(x,t)$, we
can cast Eq.~(\ref{1-d-std}) in a form that uses only first time derivatives:
\begin{equation}
{\partial \phi \over \partial t}  =  \Pi \label{eq:phi}
\end{equation}
\begin{equation}
{\partial \Pi \over \partial t} =  {\partial^2 \phi \over \partial x^2}.  \label{eq:pi}
\end{equation}
In this section we examine the system of equations~(\ref{eq:phi})--(\ref{eq:pi})
to understand the interface conditions needed for smooth propagation of
waves across mesh refinement boundaries.  In later sections, these
conditions are applied to nonlinear and multidimensional wave equations.

\subsection{Discretization}
\label{discretiz}

For the spatial discretization of equations~(\ref{eq:phi})--(\ref{eq:pi}), 
we take the data to be defined at the centers of the spatial grid cells
and use standard $O(\Delta x)^2$ centered spatial differences \cite{numrec}.
To advance this system of ordinary differential equations in time 
we use an $O(\Delta t)^2$ iterative method first suggested by M.~Choptuik 
(see Ref.~\cite{ICN}). 
In the numerical relativity literature, this explicit update scheme is refered to as 
``iterated Crank--Nicholson''.
Each iteration has the form
\begin{equation}
\phi^{n+1}_{i} = \phi^{n}_{i} + \Delta t\; \overline\Pi_{i} \label{phi-half}
\end{equation}
\begin{eqnarray}
\Pi^{n+1}_{i} & = & \Pi^{n}_{i}  +  \frac{\Delta t}{(\Delta x)^2}
(\overline\phi_{i+1} - 2 \overline\phi_{i} + \overline\phi_{i-1}) \nonumber \\
   & = & \Pi^{n}_{i} + \frac{\Delta t}{(\Delta x)^2} F(\overline\phi),
\label{pi-half}
\end{eqnarray}
where we use $i$ to label the spatial grid, $n$ to label the time steps, and
$\overline\phi_i$, $\overline\Pi_i$ to indicate intermediate values calculated during the
iteration process. Note that the familiar Crank--Nicholson algorithm is obtained by 
setting $\overline\phi_i$ and $\overline\Pi_i$ equal to their time averages, 
$(\phi^{n+1}_i + \phi^n_i)/2$ and $(\Pi^{n+1}_i + \Pi^n_i)/2$, respectively.

For two iterations, the specific steps are as follows.  Begin by applying the discretization 
(4)--(5) with $\overline\phi = \phi^n$, $\overline\Pi = \Pi^n$ to calculate 
a first approximation to 
$\phi^{n+1}$ and $\Pi^{n+1}$:
\begin{equation}
^{(1)}\phi_{i}^{n+1} = \phi_i^n + \Delta t\; \Pi_i^n    
\label{phi-tilde-1}
\end{equation}
\begin{equation}
^{(1)}\Pi_i^{n+1} = \Pi_i^n + \frac{\Delta t}{(\Delta x)^2} F(\phi^n).  
\label{pi-tilde-1}
\end{equation}
Average these new values with those at the starting time level $n$ 
to get new values for  $\overline\phi$ and $\overline\Pi$:
\begin{equation}
^{(1)}\overline\phi_{i} = \half (^{(1)}\phi_i^{n+1} + \phi_i^n)
\label{phi-bar-1}
\end{equation}
\begin{equation}
^{(1)}\overline\Pi_{i} = \half (^{(1)}\Pi_i^{n+1} + \Pi_i^n).
\label{pi-bar-1}
\end{equation}
Now perform a second iteration.  Again applying (4)--(5) we find a second approximation to 
$\phi^{n+1}$ and $\Pi^{n+1}$:
\begin{equation}
^{(2)}\phi_{i}^{n+1} = \phi_i^n + \Delta t\; ^{(1)}\overline\Pi_i   
\label{phi-tilde-2}
\end{equation}
\begin{equation}
^{(2)}\Pi_i^{n+1} = \Pi_i^n + \frac{\Delta t}{(\Delta x)^2} F(^{(1)}\overline\phi).  
\label{pi-tilde-2}
\end{equation}
Averaging again with the values at level $n$ yields
\begin{equation}
^{(2)}\overline\phi_{i} = \half (^{(2)}\phi_i^{n+1} + \phi_i^n)
\label{phi-bar-2}
\end{equation}
\begin{equation}
^{(2)}\overline\Pi_{i} = \half (^{(2)}\Pi_i^{n+1} + \Pi_i^n).
\label{pi-bar-2}
\end{equation}
A final update is carried out using these twice-iterated values:
\begin{equation}
\phi_{i}^{n+1} = \phi_i^n + \Delta t\; ^{(2)}\overline\Pi_i  
\label{phi-n+1}
\end{equation}
\begin{equation}
\Pi_i^{n+1} = \Pi_i^n + \frac{\Delta t}{\Delta x^2} F(^{(2)}\overline\phi).  
\label{pi-n+1}
\end{equation}
Clearly this algorithm can be carried out for any number of iterations. In the 
formal limit of an infinite number of iterations, it yields the usual 
Crank--Nicholson scheme. However, a von Neumann stability analysis shows that 
this iterative scheme is stable only when the number of iterations equals 
2, 3, 6, 7, 10, 11, {\it etc}, and the Courant condition $\Delta t \le \Delta x$ 
is satisfied. This was shown by Teukolsky \cite{ICN} for the advection equation, but 
the conclusion holds as well for the wave equation in the form (2)--(3). Furthermore, 
the accuracy of the iterative scheme is second order for any number of iterations. 
We must carry out at least two iterations for stability, but continuing 
beyond two iterations does not reduce the truncation error. 
In this paper we follow the common current 
practice in numerical relativity and carry out precisely two iterations for our tests.

\subsection{Evolutions on a Uniform Grid}
\label{1d-unigrid}

We first carried out uniform grid, or unigrid, evolutions of the discretized
wave equation to provide a basis for
comparison with mesh refinement runs.  The initial data for $\phi$
is taken to be a Gaussian wavepacket,
\begin{equation}
\phi(x,t=0) = A \; \e^{-x^2/\sigma^2}, \qquad \Pi(x, t=0) = 0 ,
\label{init-data}
\end{equation}
with $A = 1$ and $\sigma = 0.25$. The spatial domain
extends from $x = - 4$ to $x = + 4$.  Time evolution of this data produces two
packets traveling with velocity $v = \pm 1$, each having amplitude $A = 0.5$ and 
the same value of $\sigma$ as the original packet.  Here we will
consider only the packet traveling to the right, in the region $0 \le x \le 4$.

\begin{figure}[p]
\includegraphics[trim = 40 40 0 0, scale = 0.8]{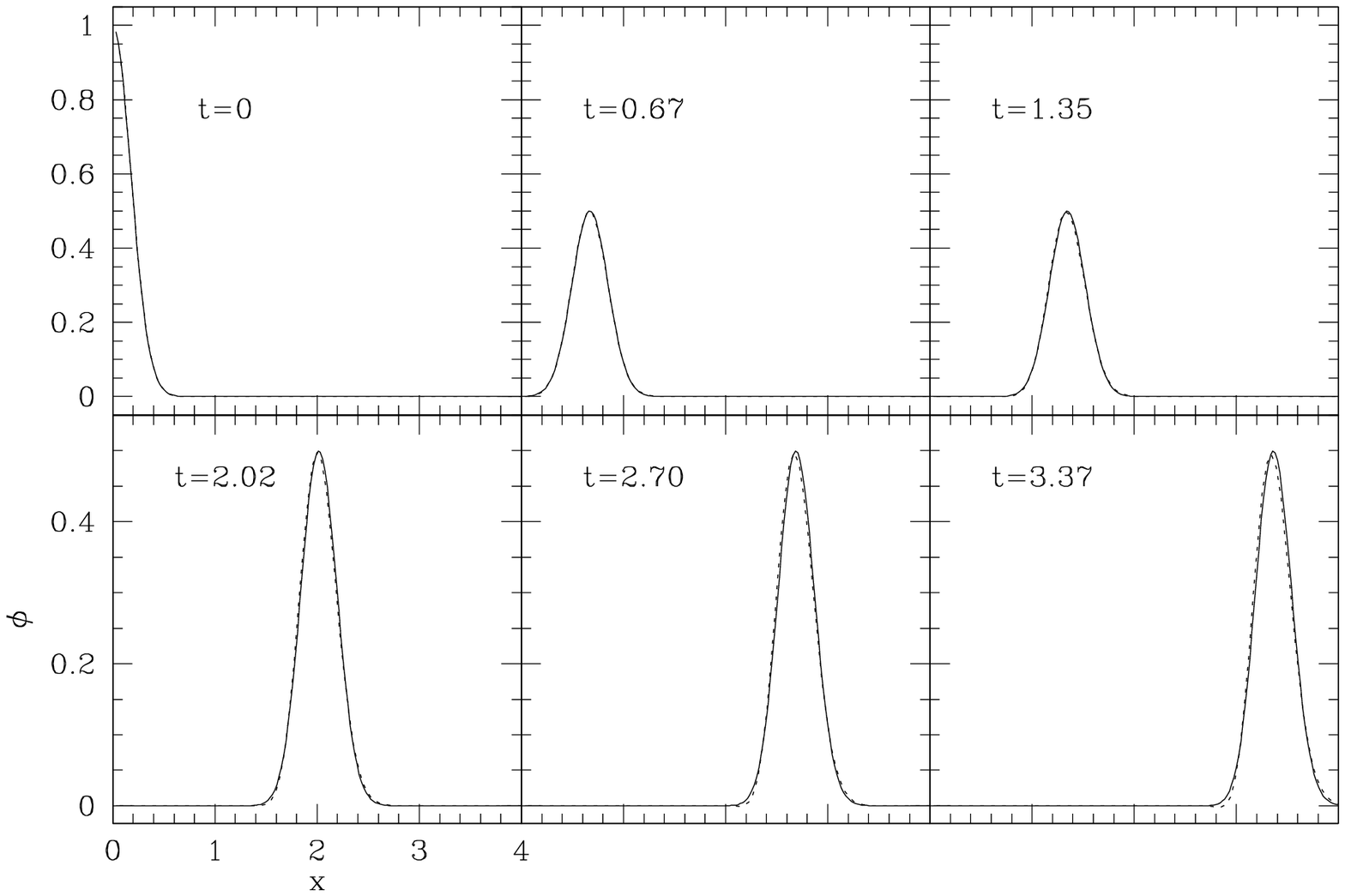}
\caption{The evolution of a Gaussian wavepacket on a uniform grid
according to the linear 1--D
wave equation is
shown for two
different resolutions: $H = \Delta x = 0.045$ (dotted line) and $h = H/2$ (solid
line).}
\label{uni-phi}
\end{figure}

Figure~\ref{uni-phi} shows the evolution of this packet for two
different resolutions. The coarser resolution is
given by $H = \Delta x = 0.045$ (dotted line), which has
$\sim 10$ zones across the width of the packet at half its maximum
amplitude.  The solid line shows resolution $h = H/2 = 0.0225$. 
The time step is chosen to be $\Delta t = \Delta x / 4$ for a given spatial
resolution, $\Delta x$.
In the last few panels 
of Fig.~\ref{uni-phi} one can see a slight separation between the two curves. This is 
primarily due to numerical dispersion, which causes the phase velocities to deviate 
from unity. The phase velocity for a monochromatic wave propogating on a discrete, uniform 
grid is calculated in the Appendix, with the result displayed in Eq.~(\ref{phase-veloc}). 
According to this formula we expect the pulse (which has wavelength $\sim 1$) to propagate 
with speed $\sim 0.999$ on the fine grid and speed $\sim 0.996$ on the coarse grid. 
This translates into a separation between the two pulses of about $0.01$ at time 
$t = 3.37$, which is the approximate separation seen  in the last panel of 
Fig.~\ref{uni-phi}.

\begin{figure}[!ht]
\includegraphics[trim = 40 40 0 0, scale = 0.8]{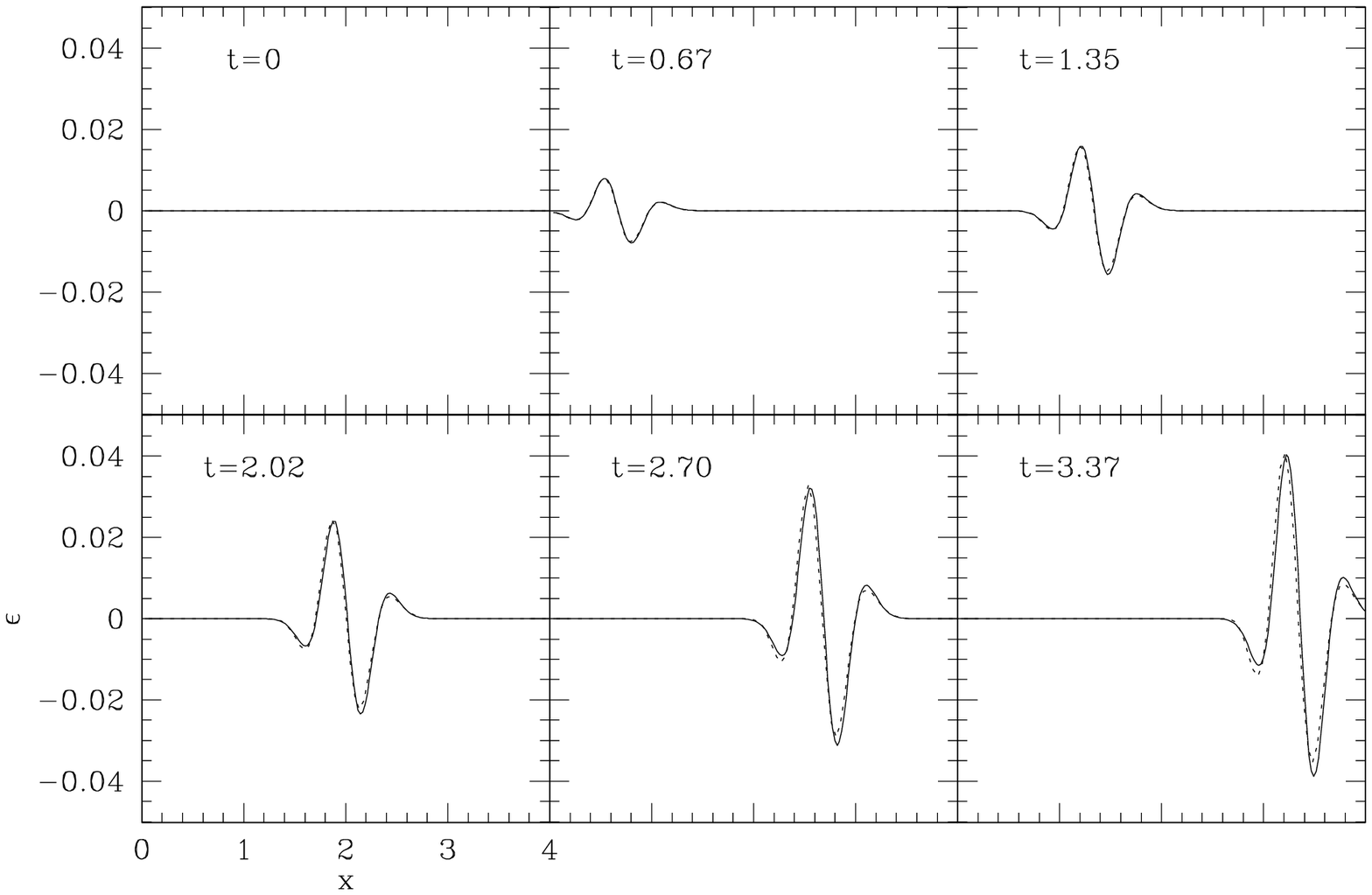}
\caption{The evolution of the absolute errors $\epsilon$ for the two
runs in
Fig.~\ref{uni-phi} is shown.  The dotted line shows $\epsilon_{\mathrm{H}}$ and 
the solid line $4 \times \epsilon_{\mathrm{h}}$.}
\label{uni-eps}
\end{figure}

The time evolution of the absolute errors $\epsilon \equiv
|\phi_{\mathrm{analytic}} - \phi_{\mathrm{numerical}}|$  is shown in
Fig.~\ref{uni-eps}.  The dotted line shows the errors
$\epsilon_{\mathrm{H}}$
for the coarse resolution $H$, and the solid line is $4 \times
\epsilon_{\mathrm{h}}$.  Inspection of Fig.~\ref{uni-eps} shows
that the two curves are nearly identical, demonstrating the second--order
convergence of these runs. Note that the errors are approximately antisymmetric about the 
location of the pulse center. This is because the dominant source of numerical error is 
dispersion, which has the principle effect of shifting 
each wave pulse relative to the exact solution. 

\section{Implementation of Mesh Refinement}
\label{descr_paramesh}

We use the Paramesh package \cite{paramesh}
 to implement the mesh refinement and parallelization in
our codes.  All of our codes use cell--centered data. 
Paramesh works on logically Cartesian, or structured, grids and carries out
mesh refinement on grid blocks.  
The underlying mesh refinement technique is similar to that of Ref. \cite{deZP},
 in which grid blocks are bisected in each coordinate direction when refinement
is needed.  The grid blocks all have the same logical structure, with $nxb$ zones in
the $x-$direction, and similarly for $nyb$ and $nzb$. Thus, 
refinement of a block in 1-D yields two child blocks, each having $nxb$ zones but with zone
sizes a factor of two smaller than in the parent block. When needed, refinement can
continue on the child blocks, with the restriction that the 
grid spacing can change only by a factor of two, or one refinement level, at any location
in the spatial domain.  
Each grid block is surrounded by a number of guard cell layers that are used in 
computing finite difference spatial derivatives near the block's boundary. These guard 
cells must be filled using data from the interior cells of the given block and the 
adjacent block.

\begin{figure}[!hb]
\includegraphics{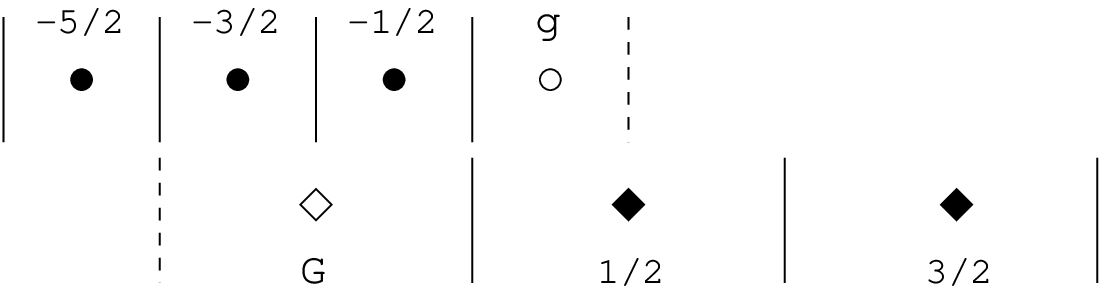}
\caption{An interlevel boundary between a coarse and fine grid in 1-D is shown.  
The coarse grid data points are marked with filled diamonds and positive half 
integers, the fine grid data points are marked with filled circles and negative 
half integers. The coarse and fine guard cells are marked with the corresponding
open symbols and are denoted by ``G'' and ``g,'' 
respectively.}
\label{fig:guard_cell}
\end{figure}

Figure~\ref{fig:guard_cell} shows a section of a 1-D grid in the vicinity of an
interlevel boundary between two neighboring grid blocks. The fine grid covers the left 
half of the 1-D space, with cell--centered grid points labeled $-1/2$, $-3/2$, {\it etc}. 
The coarse 
grid covers the right half with cell--centered grid points labeled $1/2$, $3/2$, {\it etc}. The 
fine and coarse blocks are offset from one another for clarity of presentation.
One layer of guard cells is shown, with ``G'' marking the coarse grid guard cell and ``g''
the fine grid guard cell. These guard cells are filled with data from neighboring
blocks or, if the block forms part of the edge of the computational domain, from
appropriate outer boundary conditions.

Paramesh can be used in applications requiring AMR, FMR, or a combination of these.  It 
handles the creation of grid blocks, and builds and
maintains the data structures needed to track the spatial relationships between blocks.
It takes care of all inter-block communications and keeps track of physical boundaries
on which particular conditions are set, guaranteeing that the child blocks inherit
this information from the parent blocks.  In a parallel environment, Paramesh distributes
the blocks among the available processors to achieve load balance, maximize block locality,
and minimize inter-processor communications.

For the work described in this paper, we are using FMR.  
For simplicity, we use the same timestep, chosen for stability on
the finest grid, over the entire computational domain.  At the mesh refinement
boundaries, we use a single layer of guard cells as shown in Fig.~\ref{fig:guard_cell};
special attention is paid to
the restriction (transfer of data from fine to coarse grids) and prolongation (coarse
to fine) operations used to set the data in these guard cells, as discussed in the next
subsection.

\section{Linear Wave Equation in 1-D: Evolutions with Fixed Mesh Refinement}
\label{sec:fmr}

We now carry out evolutions of 1-D linear waves that encounter a 
change in the grid resolution at a fixed location.  For the gravitational wave
applications in which we are interested, waves will be
generated in a finely resolved region and then travel out into more coarsely
resolved regions.  We thus start our initial wave packet, given by Eq.~(\ref{init-data}),
in a region of fine resolution $h = 0.0225$ around the origin.  
The spatial domain is again
$-4 \le x \le 4$.  As before, the initial wave packet splits into two identical packets
traveling in opposite directions.  Each of these packets then encounters a fixed
refinement boundary, located at $x = \pm 2.1$, and crosses into a region of coarser
resolution $H = 2h$.  In the following discussions, we focus only on the region
$x \ge 0$.

\begin{figure}[!ht]
\includegraphics[trim = 40 40 0 0, scale = 0.8]{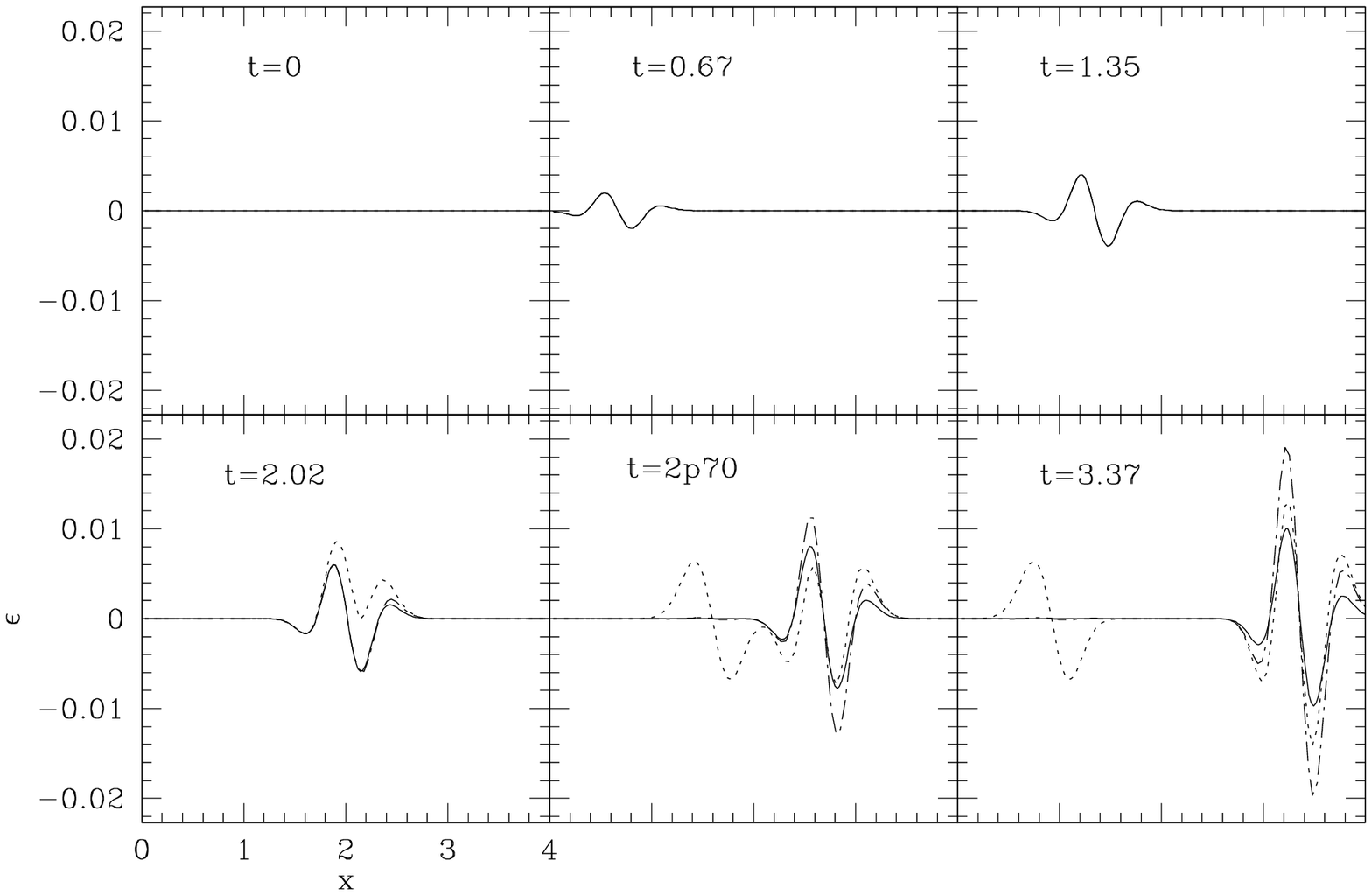}
\caption{The evolution of the absolute errors $\epsilon$ is shown for a Gaussian
packet crossing a fixed refinement boundary at $x = 2.1$ with linear (dotted line)
and quadratic (dashed line) GCF.  The solid line shows $\epsilon$ for
a unigrid run at the fine grid resolution.}
\label{eps-fmr}
\end{figure}

We first use the default Paramesh linear interpolation to set the value of the data in 
the guard cells on both the coarse and fine
grids. With this prescription for guard cell filling, the coarse grid guard cell 
value of any function $f$ is given by linear interpolation, 
\begin{equation}
	f_{\mathrm G} =\frac{1}{2}(f_{-3/2} + f_{-1/2})  .
\label{fG-lin}
\end{equation}
The value of $f$ in the fine grid guard cell ``g'' is then given by 
a linear interpolation using coarse grid values, 
$f_{\mathrm g} = ( f_{\mathrm G} + 3f_{1/2} )/4$. Combined with Eq.~(\ref{fG-lin}), this 
gives
\begin{equation}
	f_{\mathrm g} = \frac{1}{8}( f_{-3/2} + f_{-1/2} + 6f_{1/2} )  .
\label{fg-lin}
\end{equation}
Note that this guard cell filling (GCF) procedure uses the points $f_{\mathrm G}$
and $f_{1/2}$ on the coarse grid to obtain $f_{\mathrm g}$; this is in contrast to
the direct approach, which uses the nearest points $f_{-1/2}$ and $f_{1/2}$ 
({\em cf.} Eq.~(\ref{fg-direct})).  
The prescription (\ref{fG-lin})--(\ref{fg-lin}) for GCF has errors 
of order $h^2$ and is the default linear GCF method in Paramesh.  We will refer to
this procedure as {\em linear GCF} in this paper.
The results of using linear GCF are displayed in Fig.~\ref{eps-fmr}, which shows
the time evolution of the absolute errors $\epsilon$.  The dotted line shows the run
with linear interpolation at the interface boundary, and the solid line the results of
a unigrid run at the fine grid resolution.  As the packet passes through this boundary,
a reflected wave is generated propagating to the left.  The transmitted wave continues
traveling to the right into the coarse grid region.

In large scale simulations of the Einstein equations with several levels of refinement,
such spurious reflected waves can seriously degrade the quality of the results.  Globally
increasing the resolution until the reflected waves reach acceptably small amplitudes
is generally not possible in 3-D.  We thus need a better way to control the behavior
of the signals crossing the interfaces.

To this end, we implemented direct quadratic interpolation ({\em i.e.}, using the nearest
3 data points)
to set the data in the coarse and
fine grid guard cells.  Refer again to Fig.~\ref{fig:guard_cell}.  For the fine grid
guardcell ``g'', quadratic interpolation yields \cite{numrec}
\begin{equation}
f_{\mathrm g} = \frac{1}{15}(- 3 f_{-3/2} + 10f_{-1/2} + 8 f_{1/2} ).
\label{fg-quad}
\end{equation}
The coarse grid guard cell ``G'' is filled by matching first derivatives across the 
interface, 
\begin{equation}
\frac{f_{1/2} - f_{\mathrm G}}{H} = \frac{f_{\mathrm g} - f_{-1/2}}{h},
\label{flux-match}
\end{equation}
where $H = 2h$. This step, which ensures that the solution is smooth across the interface, 
can be viewed as ``flux matching'' where the gradient of $f$ plays the role of the flux. 
By combining the derivative matching condition with the formula for $f_{\mathrm g}$ we find 
\begin{equation}
f_{\mathrm G} = \frac{1}{15}( 6f_{-3/2} + 10 f_{-1/2} - f_{1/2} ).
\label{fG-quad}
\end{equation}
This same result for $f_{\mathrm G}$ can be obtained by direct quadratic interpolation. 
These formulae for GCF have errors of order $h^3$. 

The absolute errors obtained when using quadratic interpolation are shown as the dashed
line in Fig.~\ref{eps-fmr}.  Note that the reflected wave has been greatly reduced.
Additional simulations, in which the size of the zones is everywhere decreased 
by successive factors of two, show that with quadratic GCF the code is second--order 
convergent. On the other hand, with linear GCF, the reflected pulse is first--order 
convergent. The transmitted pulse also aquires first--order errors at the interface 
with linear GCF.  
As the transmitted wave propagates through the coarse grid region, second--order errors due to 
dispersion and dissipation eventually dominate over the first--order errors introduced at the 
interface. At that point, the transmitted pulse can appear second--order convergent. 

We also conducted tests using a one--dimensional periodic domain consisting of 
$20\%$ fine grid and $80\%$ coarse grid. A wave pulse was allowed to cycle through 
the domain multiple times. These tests clearly show that with quadratic guard cell filling, 
but not with linear guard cell filling,  
the code is second--order convergent. We also used this test code to check the stability 
of the interface conditions. After thousands of cycles of the wave pulse through the refined 
region, there were no signs of instability with either linear or quadratic guard cell filling. 

\begin{figure}[!htb]
\includegraphics{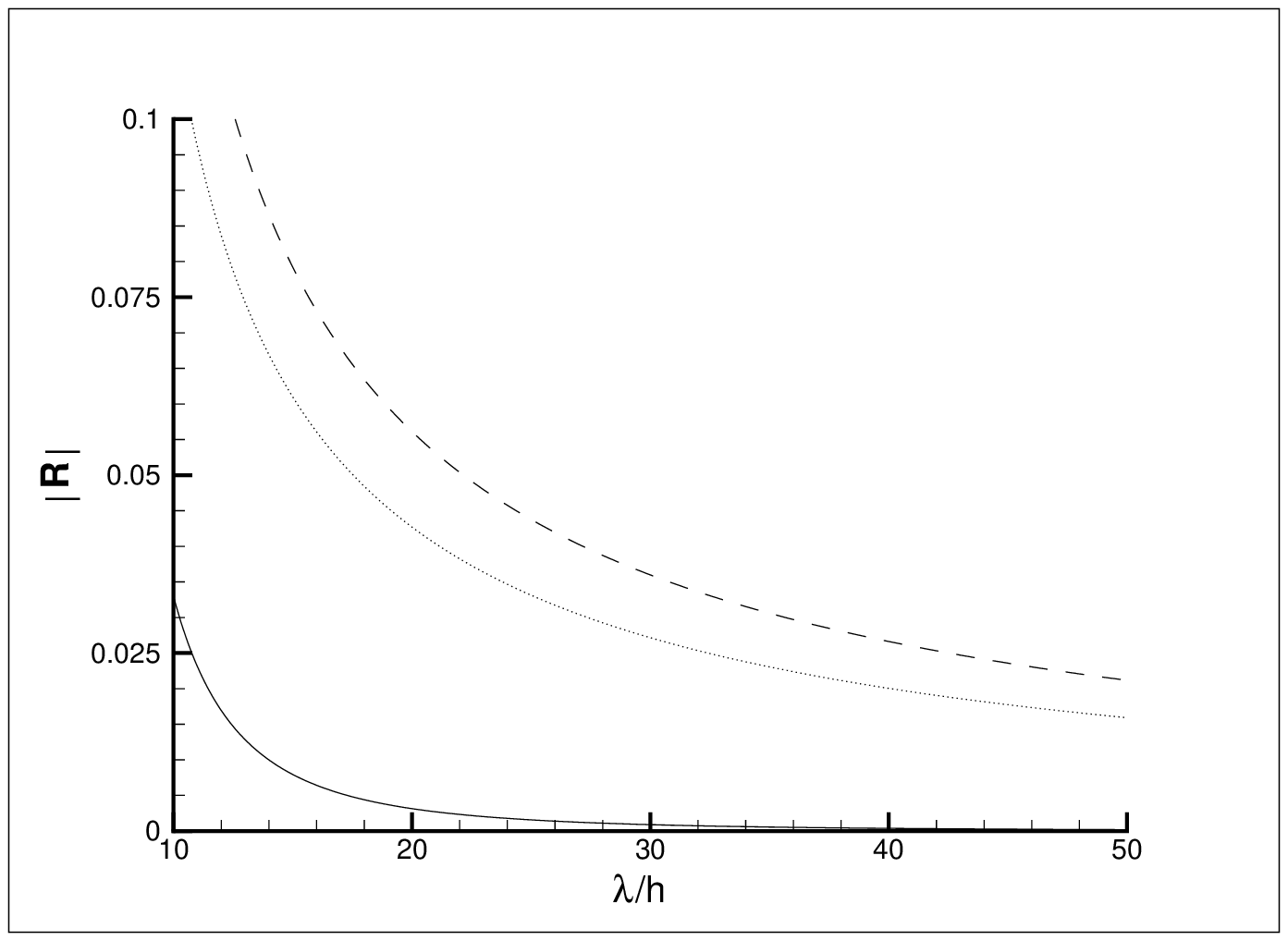}
\caption{Absolute values of the reflection coefficient $|{\bf R}|$ for linear 
(dashed curve), quadratic (solid curve), and direct linear 
(dotted curve) GCF for a wave with wavelength $\lambda$ crossing a single fixed
mesh boundary.  The resolution of the fine grid is $h$.}
\label{R_compare}
\end{figure}

In the appendix we present a detailed analytic treatment of wave propagation across
mesh refinement boundaries that complements our numerical experiments.  There we
compute the reflection coefficient ${\bf R}$ and transmission 
coefficient ${\bf T}$ for a 
monochromatic (single frequency) wave traveling on a grid with fixed mesh refinement, 
for various methods of GCF. The wave travels from a fine grid region with resolution
$h$ into a coarse grid region with resolution $2h$.
 Figure \ref{R_compare} shows the absolute 
value of the reflection coefficient $|{\bf R}|$ for 
linear GCF (\ref{fG-lin})--(\ref{fg-lin}) (dashed curve) and quadratic 
GCF (\ref{fG-quad})--(\ref{fg-quad}) (solid curve). 
The dotted curve shows the results for direct linear interpolation, defined by 
\begin{equation}
f_{\mathrm G} = \frac{1}{2}(f_{-3/2} + f_{-1/2}).
\label{fG-direct}
\end{equation}
for the coarse grid guard cell and 
\begin{equation}
f_{\mathrm g} = \frac{1}{3}(f_{-1/2} + 2 f_{1/2}).
\label{fg-direct}
\end{equation}
for the fine grid guard cell. Direct linear interpolation, like the default
linear
GCF in Paramesh, has errors of order $h^2$. 
The curves of Fig.~\ref{R_compare} are plotted as functions of 
the wavelength in the fine grid region divided by the fine grid cell size $h$. 
Equivalently, we can interpret the horizontal--axis values as the number of fine grid 
points per wavelength. 

For our 1-D wave equation tests, the Gaussian packet behaves 
roughly like a wave of wavelength $\lambda \sim 1$. With $h = 0.0225$, this 
corresponds to about $\lambda/h = 44$ fine grid points per wavelength. From 
Fig.~\ref{R_compare} we see that the reflection coefficient for linear interpolation 
is about $|{\bf R}| = 0.02$  while that for 
quadratic GCF is $|{\bf R}| = 0.0003$. With an incident pulse amplitude of $0.5$, 
we expect a reflected wave amplitude of about $0.01$ for linear GCF and 
less than $0.0002$ for quadratic GCF. 
This reflected pulse for the linear case is clearly seen in Fig.~\ref{eps-fmr}. 

The importance of minimizing spurious reflections from grid interfaces has been 
emphasized above. It is equally important to minimize the distortion of waves 
that pass through a grid interface. The errors in the transmitted wave pulse for 
linear and quadratic GCF are shown 
in the region $x > 2.1$ of the last few panels of Fig.~\ref{eps-fmr}. Note that the 
errors for quadratic GCF are actually larger than the errors for 
linear GCF. This surprising result is explained as follows. 
Observe that the errors for the two fixed mesh refinement simulations, as well as 
for the unigrid run (solid curve), are approximately antisymmetric about the 
pulse center. The errors in each case, as in the unigrid 
tests discussed in Section~\ref{lin-1d}, are primarily due to dispersion. Dispersion causes 
the wave pulses to fall behind the exact solution during propagation, giving 
rise to the errors shown in Fig.~\ref{eps-fmr}. This effect is greater for the two runs with 
fixed mesh refinement because, beyond $x=2.1$, the grid resolution is lower than for the 
unigrid run. However, with 
mesh refinement, the transmitted pulse will also suffer a phase error which has the 
effect of artificially shifting the pulse along the $x$--axis. In the case of linear
GCF, there is a relatively large positive phase error in the transmitted wave.
This phase shift partially compensates for the negative shift caused by dispersion. 
As a result the size of the largest peaks in 
the error for the transmitted wave, for the particular 
test shown in Fig.~\ref{eps-fmr}, is smaller with linear GCF than with 
quadratic GCF. 

\begin{figure}[!htb]
\includegraphics{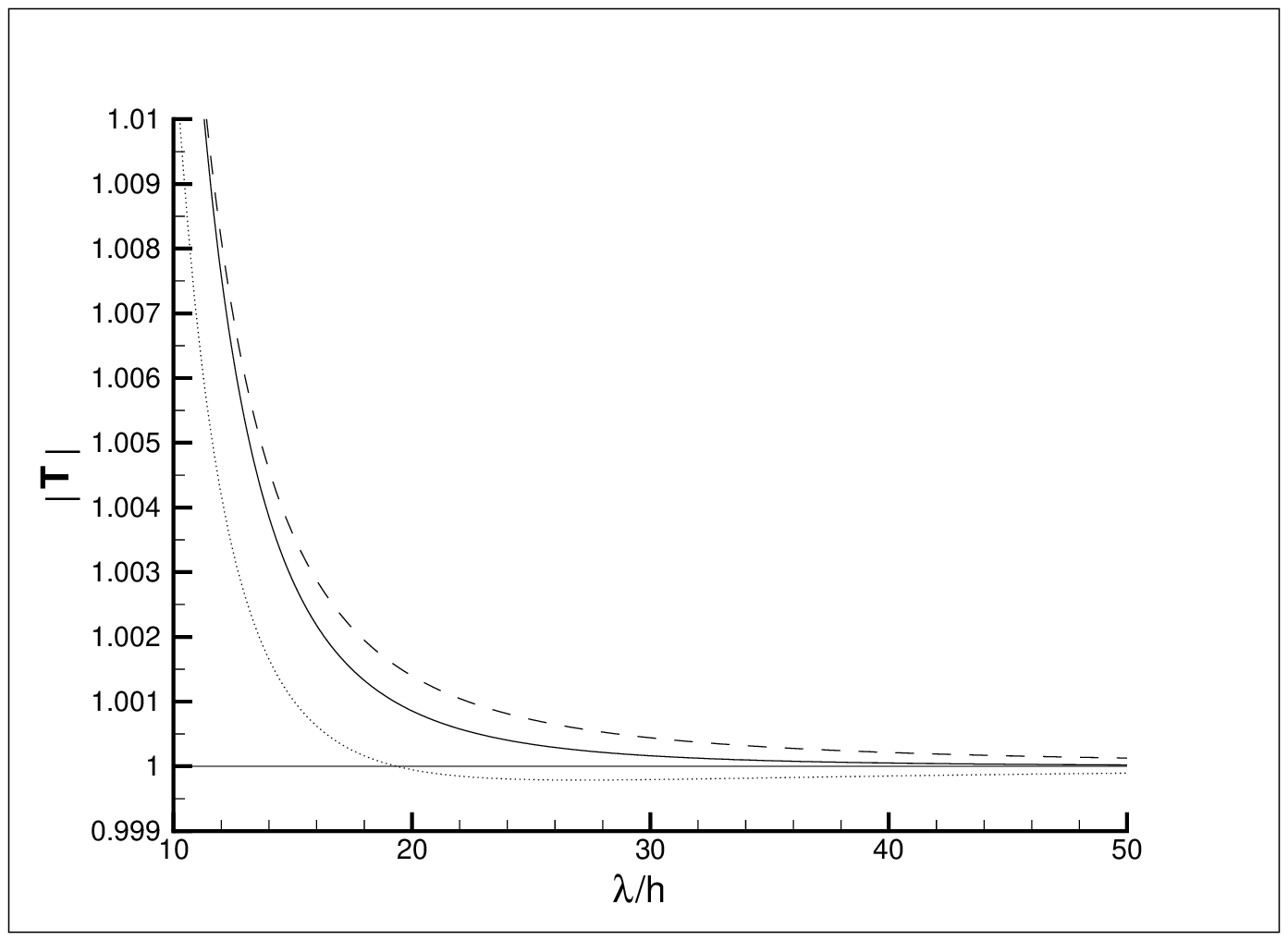}
\caption{Absolute values of the transmission coefficients for linear
(dashed line), quadratic (solid line), and direct linear (dotted) GCF.}
\label{T_compare}
\end{figure}

\begin{figure}[!htb]
\includegraphics{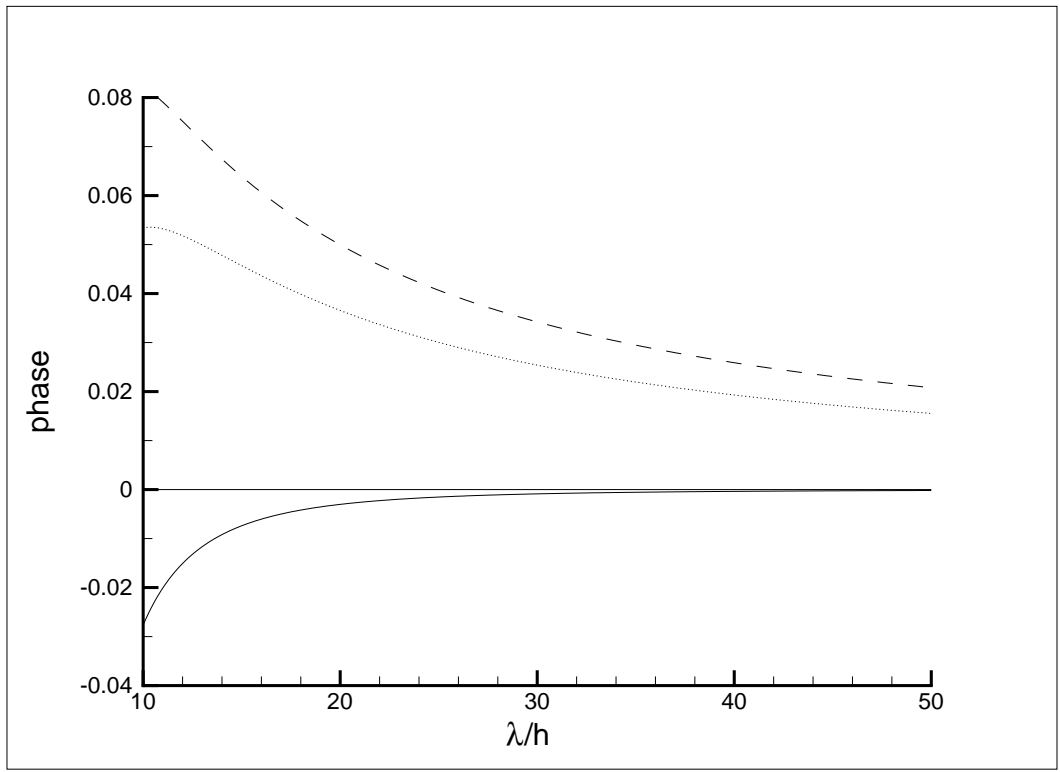}
\caption{The phase of the transmission coefficients for linear 
(dashed line), quadratic (solid line), and direct linear (dotted) GCF.}
\label{Tphase_compare}
\end{figure}

Figures \ref{T_compare} and \ref{Tphase_compare} show the absolute value of the transmission 
coefficient $|{\bf T}|$ and the phase of the transmission coefficient 
$\varphi = \arctan(\Im({\bf T})/\Re({\bf T}))$ 
for a monochromatic wave, for linear, quadratic, and direct linear interpolation. 
These graphs are obtained from the analysis in the Appendix. 
From Fig.~\ref{T_compare} it is clear that at any wavelength (any resolution) 
the error in amplitude for the transmitted wave is smaller for 
quadratic GCF than for linear GCF.\footnote{At low 
resolution, that is, for wavelengths 
less than about $28 h$, direct linear GCF has the smallest error 
for the transmitted wave amplitude. However, as discussed in the Appendix, as the 
resolution is increased $|{\bf T}|$ is much closer to $1$ for quadratic 
GCF. Also note from Fig.~\ref{Tphase_compare} that 
direct linear GCF has large phase errors for the transmitted wave.}
The dominant source of error for the transmitted wave is actually phase error, 
shown in Fig.~\ref{Tphase_compare}. The magnitude of this error for quadratic GCF 
is much smaller than that for linear GCF. For a 
wavelength of $\lambda \sim 1$, the linear
guard cell filling produces a phase shift of about $\varphi = 0.024$, while  
quadratic GCF gives a phase shift of about $\varphi = -0.00028$. For the 
tests shown in Fig.~\ref{eps-fmr}, the positive phase for linear GCF 
translates into a shift along the positive $x$--axis of about 
$\delta x = \lambda\varphi/(2\pi) \approx 0.004$. 
With quadratic GCF, the pulse is shifted in the negative direction, 
but by a much smaller amount $\delta x \approx -0.00004$. Close inspection of the data for 
the two transmitted pulses shows that they indeed have a separation of 
$\delta x \approx 0.004$. For linear GCF, this phase shift pushes the 
wave pulse forward and artificially compensates for the phase lag caused by dispersion. In 
general, there is no reason to expect the cumulative phase lag due to dispersion to be close in 
magnitude (but opposite in sign) to the phase advance caused by transmission through 
various grid interfaces. Thus, the relatively small transmission error seen in 
Fig.~\ref{eps-fmr} for linear GCF should be viewed as an accident 
of the particular example, not a generic result.

\section{Nonlinear Wave Equation in 1-D}
\label{nonlin-1d}

The next step in developing model equations to test these interface conditions 
is to add nonlinear terms similar to those found in the Einstein equations.  
This produces the following nonlinear
wave equation
\begin{equation}
{\partial^2 \phi \over \partial t^2} =  {\partial^2 \phi \over \partial x^2} 
+ d \left( {\partial \phi \over \partial t} \right)^2 + 
e  \left( {{\partial \phi \over \partial x}} \right)^2 ,
\label{1-d-nonlin}
\end{equation}
where $d$ and $e$ are arbitrary contants.  Again introducing the auxiliary variable
$\Pi(x,t)$, we get the first order system
\begin{equation}
{\partial \phi \over \partial t} = \Pi 
\label{phi-nonlin}
\end{equation}
\begin{equation}
{\partial \Pi \over \partial t} =  {\partial^2 \phi \over \partial x^2} 
+ d\; \Pi^2 + e \left( {\partial \phi \over \partial x} \right) ^2.
\label{pi-nonlin}
\end{equation}

Using the discretization introduced in \S~\ref{discretiz}, we have
\begin{equation}
\phi^{n+1}_{i} = \phi^{n}_{i} + (\Delta t)  \overline\Pi_{i} 
\label{phi-nonlin-dis}
\end{equation}
\begin{eqnarray}
\Pi^{n+1}_{i} & = & \Pi^{n}_{i} + {\Delta t \over (\Delta x)^2} 
(\overline\phi_{i+1} - 2 \overline\phi_{i} + \overline\phi_{i-1})
 + d (\Delta t) (\overline\Pi_{i} )^2    \nonumber \\
&  &  + e (\Delta t) \left( {\overline\phi_{i+1} - \overline\phi_{i-1}
\over 2 \Delta x} \right)^2 .
\label{pi-nonlin-dis}
\end{eqnarray}
Equations~(\ref{phi-nonlin-dis}) and~(\ref{pi-nonlin-dis}) are updated following the
steps given in~(\ref{phi-tilde-1})--(\ref{pi-n+1}).

We consider the case $d = -e = 1$ and set up an initial Gaussian wave packet centered
on the origin using the prescription given by Eq.~(\ref{init-data}),
with $\Pi(x,t=0) = 0$. 
This splits into two identical packets traveling in opposite directions, each
having  amplitude $A = 0.38$ and width $\sigma = 0.25$.
We use the spatial domain $-4 \le x \le 4$ and set fixed refinement
boundaries at $x = \pm 2.1$.  The fine grid 
around the origin has resolution $h = 0.0225$
and the coarse grid regions have resolution $H = 2h$. We focus on the region
$x \ge 0$.

\begin{figure}[!ht]
\includegraphics[trim = 40 40 0 0, scale = 0.8]{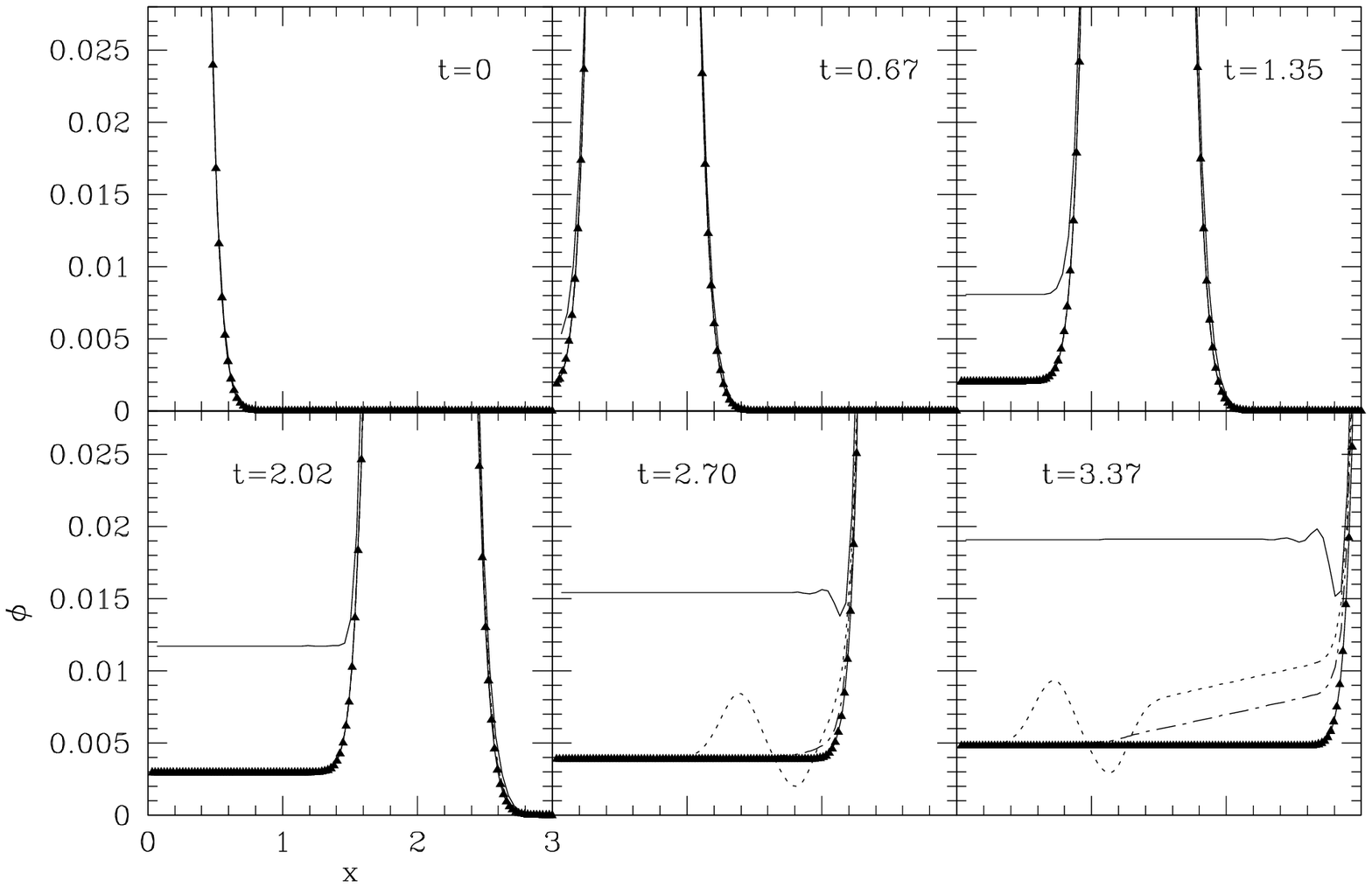}
\caption{The time evolution of the solution $\phi$ to the nonlinear 1-D wave equation
is shown.  Unigrid runs are shown at the coarse resolution $H$ (thin solid line) and
the fine resolution $h=H/2$ (thick solid line).  Runs in which the packet encounters a 
mesh refinement boundary at $x = 2.1$ are shown for linear interpolation (dotted line)
and quadratic (dashed line) GCF.}
\label{fig:1-d_nonlin}
\end{figure}

The results are shown in Figure~\ref{fig:1-d_nonlin}.  Since we do not have an analytic
solution for Eq.~(\ref{1-d-nonlin}), we display the actual solution and use unigrid
runs for comparison.  In addition, the vertical scale is chosen to zoom in on the
region around the base of the packet ({\em i.e.}, near $\phi = 0$), 
where the differences between the runs are the most
apparent. The thin solid line shows the solution for a unigrid run at the
coarse resolution $H$, and the thick solid line shows a unigrid run at the fine resolution
$h$. Runs in which the packet encounters a refinement
boundary are shown using a dotted line (linear GCF) and a dashed line
(quadratic GCF).  As we saw before, a reflected wave is generated when
the packet crosses the refinement boundary using linear GCF; these effects
are much less noticeable when using quadratic GCF.  As in the case of the 
linear wave equation, the code is second--order convergent when using quadratic GCF.

\section{Wave Equation in 2--D}
\label{2D}

As a next step, we consider the wave equation in 2--D.  The
evolution of cylindrically symmetric waves on a 2--D Cartesian mesh provides an
ideal test problem in which the signals cross mesh refinement boundaries that are,
in general, not perpendicular to their directions of propagation.

The 2--D model wave equation takes the form
\begin{equation}
{\partial^2 \phi \over \partial t^2} =  {\partial^2 \phi \over \partial x^2}
+ {\partial^2 \phi \over \partial y^2}
+ d \left( {\partial \phi \over \partial t} \right)^2
+ e_1 \left( {{\partial \phi \over \partial x}} \right)^2
+ e_2  \left( {{\partial \phi \over \partial y}} \right)^2 ,
\label{2d-eqn}
\end{equation}
where $d,e_1,e_2$ are contants.  With the auxiliary variable $\Pi(x,y,t)$, we
can write this in a form using only first--order time derivatives:
\begin{equation}
{\partial \phi \over \partial t}  =  \Pi 
\label{phi-2d}
\end{equation}
\begin{equation}
{\partial \Pi \over \partial t}  =   {\partial^2 \phi \over \partial x^2} 
+  {\partial^2 \phi \over \partial y^2}
+ d\; (\Pi)^2 + e_1 \left( {\partial \phi \over \partial x} \right)^2 
            + e_2  \left( {\partial \phi \over \partial y} \right)^2.
\label{pi-2d}
\end{equation}

Using the discretization introduced in \S~\ref{discretiz}, we have
\begin{equation}
\phi^{n+1}_{ij} = \phi^{n}_{ij}  + (\Delta t)  \overline\Pi_{ij} 
\label{phi-2d-dis}
\end{equation}
\begin{eqnarray}
 \Pi^{n+1}_{ij} & = &  \Pi^{n}_{ij} +
      {\Delta t \over (\Delta x)^2} 
(\overline\phi_{i+1,j} -2\overline\phi_{ij} + \overline\phi_{i-1,j}) 
\nonumber \\
&  &  + \; {\Delta t \over (\Delta y)^2}
(\overline\phi_{i,j+1} -2\overline\phi_{ij} + \overline\phi_{i,j-1}) 
\nonumber \\
 & & +\; d (\Delta t) ( \overline\Pi_{ij}  )^2  
  +  e_1 (\Delta t) \left( {\overline\phi_{i+1,j}-\overline\phi_{i-1,j}
 \over 2 \Delta x} \right)^2  \nonumber \\
 &  & +\;  e_2 (\Delta t) \left( {\overline\phi_{i,j+1}-\overline\phi_{i,j-1} 
 \over 2 \Delta y} \right)^2 .
\label{pi-2d-dis}
\end{eqnarray}
As before, Eqs.~(\ref{phi-2d-dis}) and~(\ref{pi-2d-dis}) are updated following the
steps given in Eqs.~(\ref{phi-tilde-1})--(\ref{pi-n+1}).

\begin{figure}[!ht]
\centerline{\includegraphics{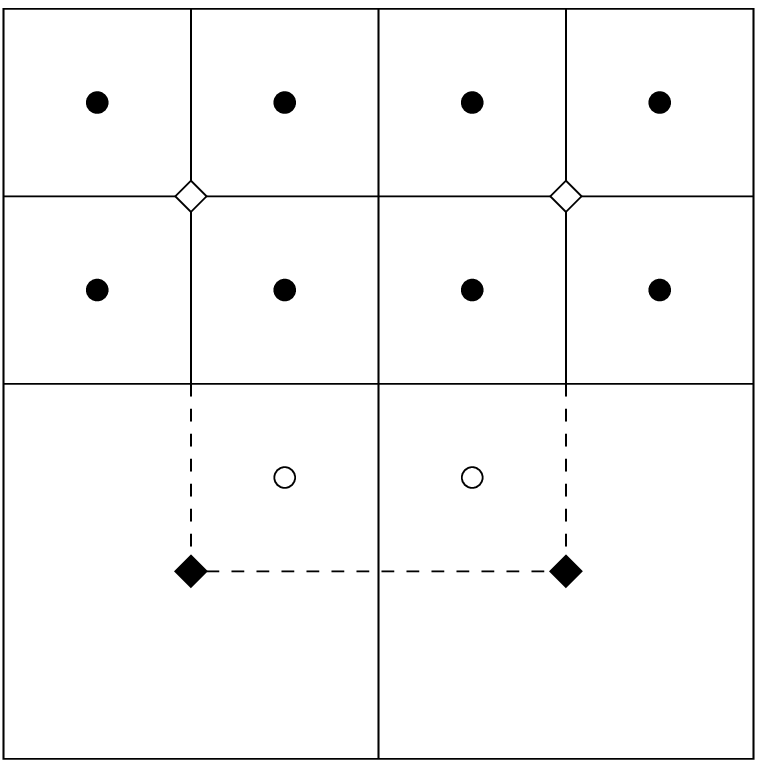}}
\caption{Paramesh default linear GCF in 2-D. Points on the coarse grid are denoted by
diamonds, and points on the fine grid by circles. First, the coarse grid guard cells (open 
diamonds) are filled by averaging the surrounding fine grid points. Then the fine grid 
guard cells (open circles) are filled by taking linear combinations of the four surrounding 
coarse grid points.}
\label{fig_pgcf}
\end{figure}

In this section we consider two types of GCF, the default Paramesh linear 
order GCF and a quadratic GCF scheme.  The linear 
GCF is depicted in Fig.~\ref{fig_pgcf}. 
First, each coarse grid guard cell 
(open diamond) is filled as a linear combination of the surrounding fine grid points (solid 
circles). The fine grid guard cells (open circles) are then filled using a linear combination 
of the surrounding coarse grid points (open and solid diamonds). 

\begin{figure}[!ht]
\centerline{\includegraphics{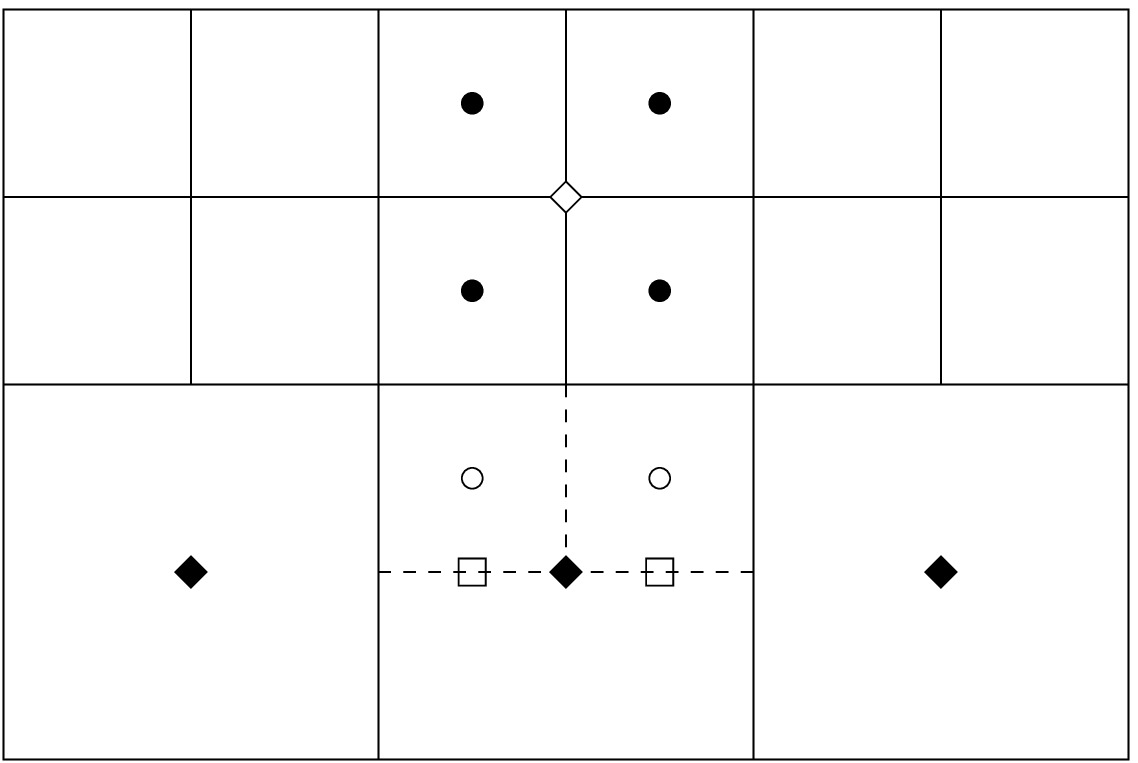}}
\caption{Quadratic GCF in 2-D. Coarse grid guard cells (open diamonds) 
are temporarily filled by averaging the surrounding fine grid points. Intermediate 
values (open boxes) 
are obtained by linear interpolation of coarse grid values (solid diamonds)
parallel to the interface. 
Fine grid guard cells (open circles) are then filled from a quadratic fit using two fine 
grid points and one intermediate value. Finally, the coarse grid guard cells are 
filled by matching the coarse and fine grid first derivatives across the interface.}
\label{fig_qgcf}
\end{figure}

Quadratic GCF is depicted in Fig.~\ref{fig_qgcf}. 
As a first step, 
the coarse grid guard cells are filled from a linear combination of the four surrounding 
fine grid guard cells. These values are only used at fine grid corners, and 
will soon be overwritten. Linear 
interpolation of the coarse grid cells (solid diamonds) 
parallel to the coarse--fine interface is used 
to compute intermediate values marked with open boxes in Fig.~\ref{fig_qgcf}. These 
intermediate values, along with the two fine grid cells (solid circles) 
directly across the interface, 
are then used to obtain a quadratic fit for the fine grid guard cells marked with open 
circles. 

Finally, as in the 1-D case, the coarse grid guard cells are filled by 
``flux matching'', that is, matching derivatives across the interface. 
Specifically, we consider the first derivative at the midpoint of 
Fig.~\ref{fig_qgcf}, that is, at the point midway between the coarse grid guard cell 
(open diamond) and the interior cell directly across the interface (closed diamond). 
Derivative matching consists in equating the first derivative computed 
from these coarse grid cells with the second--order accurate first derivative obtained 
from the four fine grid cells (open and closed circles) that surround the midpoint. 

The algorithm described here for quadratic GCF is similar to the one 
described by Martin and Cartwright \cite{martin}. The main difference is that we use linear 
interpolation of coarse grid values parallel to 
the interface to obtain intermediate values (the open 
boxes in Fig.~\ref{fig_qgcf}),  whereas 
Martin and Cartwright use quadratic interpolation. Also 
note that our algorithm can be applied without 
modification at fine grid corners, where the corner of a coarse grid block is 
surrounded by fine grid blocks. Recall that in the first step, 
coarse grid guard cells are filled by linear restriction from the fine grid. This allows 
the interpolation parallel to the interfaces to be carried out without the use 
of one--sided extrapolation. Finally, we point out that GCF at fine 
grid corners
is ambiguous, since there are different ways to deal with them;
either of the two coarse--fine interfaces that intersect 
at the corner can be used 
or intepolation using a stencil diagonal to the interfaces can also be used. 
Note that only mixed derivatives are affected by the corners when using centered 
differencing.
In our code we do not treat the corners as special. At a corner our code naturally 
selects one of the two interfaces and carries out a linear interpolation parallel to 
that face to obtain intermediate values.

The initial data for our tests is taken to be a cylindrically symmetric wavepacket 
centered on the origin, with
\begin{equation}
 \phi(x,y,t=0)=A e^{-(x^2+y^2)/\sigma^2}
\label{2d-packet}
\end{equation}
and $\Pi(x,y,t=0) = 0$.  We choose the amplitude $A = 1$,
and the width of the pulse by $\sigma = 0.25$.
Quadrant symmetry is imposed by using mirror-symmetry boundary conditions
along $x = 0$ and $y = 0$.  The computational domain then covers the region
$0 \le x \le 4.3125$ and $0 \le y \le 4.3125$.
This packet is initially confined to a fine grid region of resolution 
$h = 0.0225$.  As the
packet expands, the wavefront crosses a fixed mesh refinement boundary 
into a region of coarser resolution $H = 2h$.

\begin{figure}[!ht]
\includegraphics[trim = 40 40 0 0, scale = 0.8]{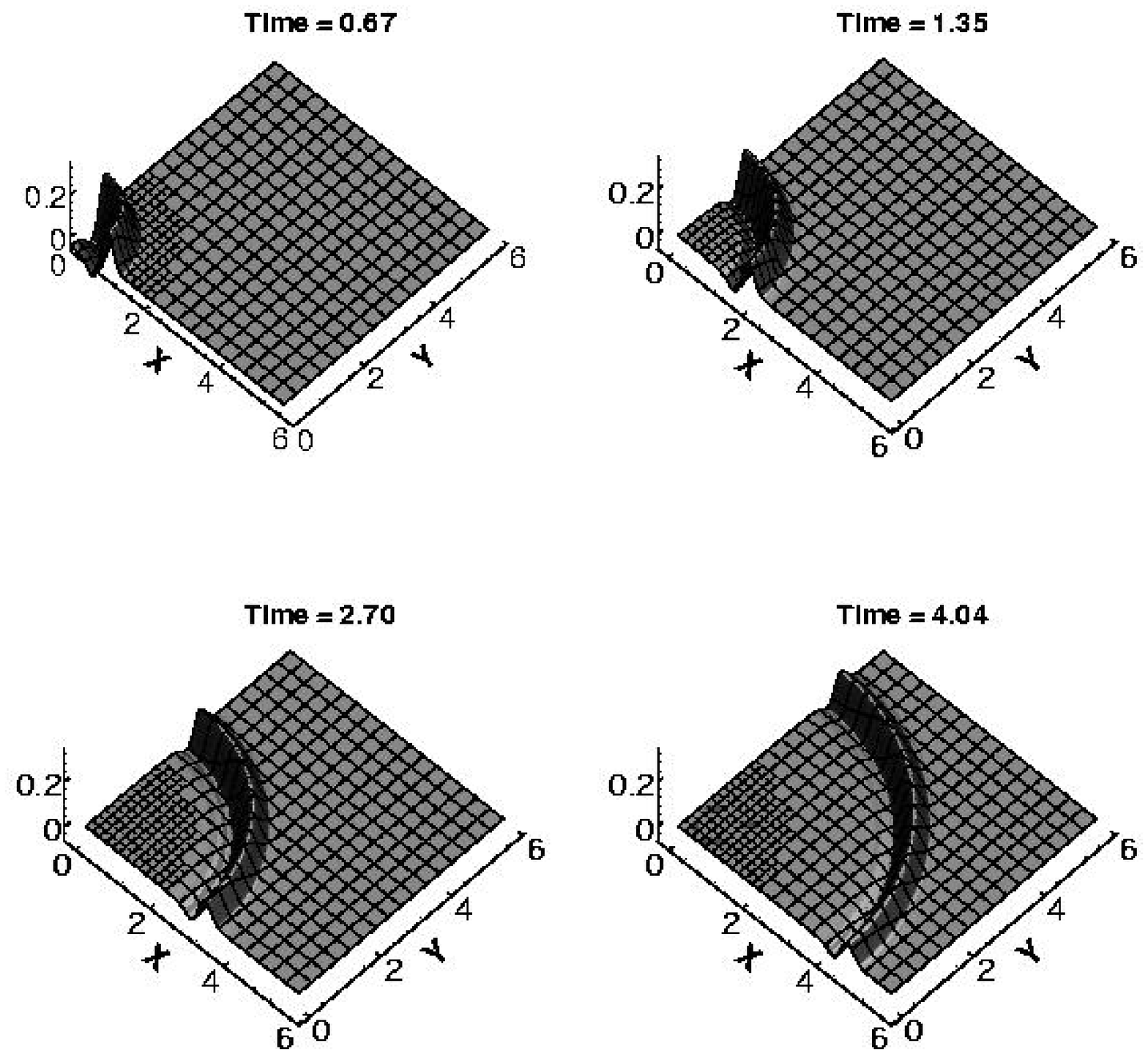}
\caption {The evolution of $\phi$ with a linear 2-D wave equation is shown at 4 consecutive
times for a run using quadratic GCF to set the data in the guard cells. Each of
the grid blocks shown has $8 \times 8$ zones. }
\label{2d-lin}
\end{figure}

Setting $d = e_1 = e_2 = 0$ in Eqs.~(\ref{phi-2d}) and~(\ref{pi-2d}) allows this 
packet to evolve under a linear equation. 
Figure~\ref{2d-lin} shows the results of using quadratic GCF to set 
the values of the guard cell data.  Here, $\phi$ is shown at four 
consecutive times.  The expanding
wavefront encounters mesh refinement boundaries at $x = 2.1$ along the $x-$axis and at $y = 2.1$
along the $y-$axis.  Note that the wave passes smoothly across the interface.

\begin{figure}[!ht]
\centerline{\epsfxsize=16.0cm\epsfxsize=16.0cm\epsffile{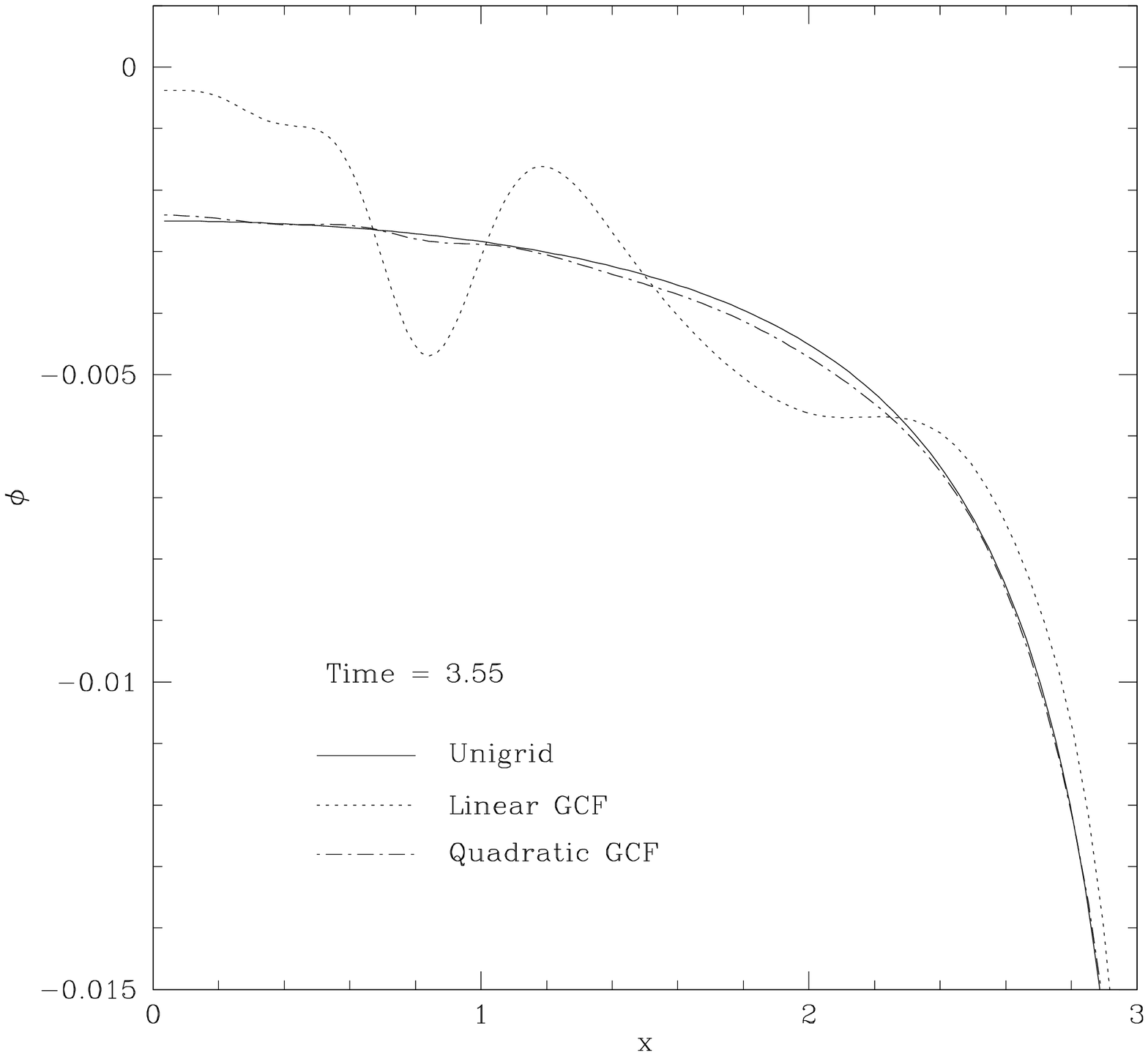}}
\caption{$\phi$ is shown along a portion of the $x-$axis for the 2-D linear wave equation. 
The solid line shows the results of a unigrid run at the fine grid resolution $h=0.0225$, 
the dotted line the results of a run with linear GCF, and the
dashed line a run with quadratic GCF.}
\label{2d-compare}
\end{figure}

A comparison of unigrid and fixed refinement runs is shown in Fig.~\ref{2d-compare}.  
Here, $\phi$
is plotted along a portion of the $x-$axis at a fixed time.  The unigrid run (solid line) 
at the fine grid resolution shows the extended ``tail'' of the outgoing cylindrical
wave front.  The run with linear GCF (dotted line) shows a reflected wave 
traveling back into the fine grid region as the wave
passes through the refinement boundary.  In the run with quadratic GCF (dashed line), 
this spurious signal has been nearly eliminated.

\begin{figure}[!ht]
\includegraphics[trim = 20 20 0 0, scale = 0.75]{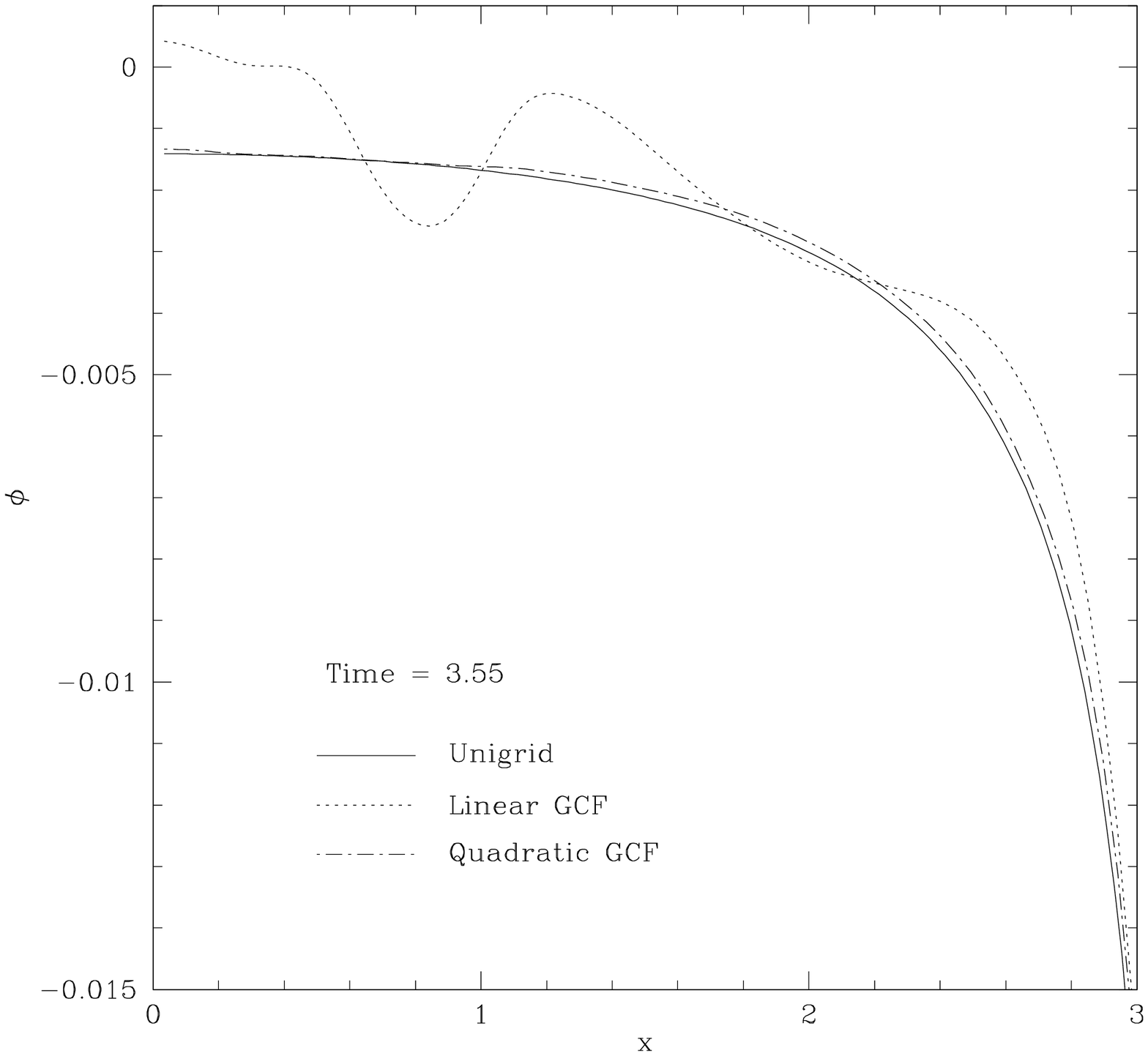}
\caption{Same as Fig.~\ref{2d-compare} except that the 2-D nonlinear wave equation is solved.}
\label{2d-nonlin-compare}
\end{figure}

Similar results are achieved when this wave packet is evolved according to a nonlinear equation,
$d = -e_1 = -e_2 = 1$.  The structure of the grid and location of the refinement boundary 
are the
same as for the 2-D linear equation.  Figure~\ref{2d-nonlin-compare} displays the results of
$\phi$ along the $x-$axis at a fixed time.  Notice that the run with linear GCF
(dotted line)
shows a significant reflected wave.  In contrast, the run with quadratic GCF
(dashed line) is close to the one with a uniform grid (solid line).  

\section{The Einstein Equations in 3--D}
\label{EE}

We are now ready to apply the techniques developed in our model equations to 
the propagation of gravitational waves in 3--D, which is governed by the
vacuum (or source--free) 
Einstein equations.  We write these equations in terms of the 
``3 + 1'' spacetime split \cite{MTW}, in which the initial data is specified
on some 3--D spacelike slice and then evolved forward in time.  Within
this framework, the metric takes the form
\begin{equation}
ds^2 = -\alpha^2 dt^2 + g_{ij} (dx^i + \beta^i dt)(dx^j + \beta^j dt).
\label{3-metric}
\end{equation}
We use units in which both the speed of light $c=1$ and the gravitational 
constant $G=1$.  Lowercase Latin letters are used to denote spatial indices, so
that $i,j = 1,2,3$.  To simplify the notation throughout this section, we use
the summation convention: if any expression has one index as a superscript and
the same index as a subscript, summation over all values that index can take is
implied \cite{schutz}.  The geometry of the given spacelike slice is described by
the 3--metric $g_{ij}$.  The lapse
function $\alpha$ governs the advance of proper time across the surface, and the
shift vector $\beta^i$ the motion of the spatial coordinates within the
hypersurface as the data is evolved forward in time. Both $\alpha$ and
$\beta^i$ are freely--specifiable functions of space and time; 
for the rest of this section,
we use the choice $\alpha = 1$ and $\beta^i  = 0$. 

In the standard ADM spacetime split \cite{MTW}, the Einstein equations 
can be written in terms of $g_{ij}$ and the extrinsic curvature of the
hypersurface $K_{ij}$, where
\begin{equation}
	K_{ij} = - \frac{1}{2} {\partial g_{ij} \over \partial t}.
\label{Kij}
\end{equation}
Following current practice in numerical relativity, we use the BSSN
formalism \cite{SN,BS} in which the Einstein equations are written in terms of
conformal
variables \{$\psi, K, \tilde{g}_{ij}, \tilde{A}_{ij}, \tilde \Gamma^{i}$\}
defined as follows:
\begin{equation}
e^{4 \psi} \equiv  {\rm det}(g_{ij})^{1/3} 
\end{equation}
\begin{equation}
\tilde{g}_{ij} \equiv  e^{-4 \psi} g_{ij} 
\end{equation}
\begin{equation}
K  \equiv g^{ij} K_{ij} 
\end{equation}
\begin{equation}
\tilde{A}_{ij} \equiv e^{-4 \psi} ( K_{ij} - {1\over 3} g_{ij} K ) 
\end{equation}
\begin{equation}
\tilde{\Gamma}^{i} \equiv - \partial_j \tilde{g}^{ij} \; .
\label{connecfun}
\end{equation}
Here $\tilde{g}^{ij}$ is the inverse  of the conformal
metric  $\tilde{g}_{ij}$.
We use the notation $\partial_{j} \equiv \partial/\partial x^j$ for
spatial derivatives.  

In terms of these conformal variables, with the gauge choices
$\alpha = 1$ and $\beta^i = 0$,
the vacuum Einstein equations become
\begin{equation}
{\partial \psi  \over \partial t}=  - {1\over6} \alpha K   
\label{EE-phidot}    
\end{equation}
\begin{equation}
{\partial \tilde{g}_{ij}  \over \partial t} = - 2 \tilde{A}_{ij}  
\end{equation}
\begin{equation}
{\partial K \over \partial t} =
    {1\over3} K^2+\tilde{A}_{ij}\tilde{A}^{ij}
\end{equation}
\begin{equation}
{\partial \tilde{A}_{ij} \over \partial t}= 
    R^{\mathrm {TF}}_{ij} +
 \tilde{A}_{ij} K - 2 \tilde{A}_{il} \tilde{A}^{l}{}_{j}
\label{EE-aijdot}
\end{equation}
\begin{equation}
{\partial \tilde{\Gamma}^{i} \over \partial t}= 
2 (  \tilde{\Gamma}^{i}{}_{jk} \tilde{A}^{kj}
            - {2\over3} \tilde{g}^{ij} \partial_j K
            + 6 \tilde{A}^{ij} \partial_j \psi ) .
\label{EE-Gammadot}
\end{equation}
Here, $\tilde{\Gamma}^{i}{}_{jk}$ are
the connection coefficients associated with $\tilde{g}_{ij}$,
defined by 
\begin{equation}
\tilde \Gamma^{k}{}_{ji} = {1\over2} \tilde{g}^{mk}(\partial_i \tilde{g}_{mj} 
+ \partial_j \tilde{g}_{mi} - \partial_m \tilde{g}_{ji}),
\label{connec}
\end{equation}
and
\begin{equation}
\tilde{A}^{ij} = \tilde{g}^{il} \tilde{g}^{jk} \tilde{A}_{lk}, \qquad 
\tilde{A}^{l}{}_{j} = \tilde{g}^{li} \tilde{A}_{ij} .
\end{equation}
The superscript ``TF'' denotes the trace-free part of a tensor, so that
$R^{\mathrm {TF}}_{ij} = R_{ij} - g_{ij} R/3$, where $R = g^{mk}R_{mk}$.
The Ricci curvature tensor $R_{ij}$ is defined by
\begin{equation}
R_{ij} = \partial_k \Gamma^{k}{}_{ij} - \partial_j \Gamma^{k}{}_{ik}
+ \Gamma^{k}{}_{mk} \Gamma^{m}{}_{ij} - \Gamma^{k}{}_{mj} \Gamma^{m}{}_{ik}.
\label{ricci}
\end{equation}

Although
the set of equations~(\ref{EE-phidot})--(\ref{EE-Gammadot}) 
is considerably more complicated than our model equations, there are
notable similarities.  In particular, the conformal metric
$\tilde g_{ij}$ plays the role of the function $\phi$, while 
$A_{ij}$ takes the role of $\Pi$.  Looking at ~(\ref{connec})
and~(\ref{ricci}), we also see that the $R^{\mathrm {TF}}_{ij}$ term
in Eq.~(\ref{EE-aijdot}) contains second spatial derivatives of
$\tilde g_{ij}$.

The lessons learned from the model equations in 1--D and 2--D can be applied 
successfully to the Einstein equations in 3--D, as we demonstrate by evolving
a weak gravitational wave.  We use the analytic
solution to the linearized Einstein equations 
found by Teukolsky \cite{teuk_lw}; since this is given in closed form, we can
then
compare the numerical results directly with this analytic solution.
We choose the even parity, $L=2$, $M=0$ solution, which is given by
\begin{eqnarray}
ds^2 & = & -dt^2
     + (1+A f_{rr}) dr^2 
     + (2Bf_{r \theta}) r dr d\theta \nonumber \\
 & & + (2Bf_{r \phi}) r \sin \theta dr d\phi
     + (1+ C f^{(1)}_{\theta \theta} + A f^{(2)}_{\theta \theta}) r^2 d\theta^2 \nonumber \\
 & & + [2(A-2C)f_{\theta \phi}]r^2 \sin \theta d\theta d\phi  \nonumber \\
 & & + (1+ C f^{(1)}_{\phi \phi} + A f^{(2)}_{\phi \phi}) r^2 \sin^2 \theta d\phi^2 .
\end{eqnarray}
Here,   
\begin{eqnarray}
A & = & 3 \left [ {F^{(2)} \over r^3} + {3F^{(1)} \over r^4}+{3F \over r^5} \right ] \\
B & = & - \left [ {F^{(3)} \over r^4} + {3F^{(2)} \over r^3}+{6 F^{(1)}\over r^4}+{6F\over r^5} \right ] \\
C & = & {1 \over 4} \left [{ F^{(4)}\over r }+{2F^{(3)} \over r^2}+{9F^{(2)} \over r^3}
  +{21 F^{(1)} \over r^4}+{21 F \over r^5} \right ] \\
F & = & F(t-r), \;\;\;\;\;  F^{(n)} \equiv \left [{d^{n} F(x) \over dx^{n}} \right ]_{x=t-r} ,
\end{eqnarray}
where $F$ is a generating function.
We use the form
\begin{eqnarray}
F (x) = {A x \over \omega^2} e^{-{x^2 / \omega^2}},
\end{eqnarray}
with two free parameters, $A$ and $\omega$. 
Here we have specified an outgoing wave solution
$F = F(t-r)$; an ingoing wave solution can be obtained by using $F = F(t+r)$.

For this even-parity, $M=0$ case, the 
angular functions $f_{ij}$ are:
\begin{eqnarray}
f_{rr}                  &=&  2 - 3\sin^2 \theta \\
f_{r\theta}             &=&  -3 \sin\theta \cos \theta \\
f_{r\phi}               &=&  0 \\
f^{(1)}_{\theta \theta} &=&  3 \sin^2 \theta  \\
f^{(2)}_{\theta \theta} &=&  -1 \\
f_{\theta \phi}         &=&  0 \\
f^{(1)}_{\phi \phi}     &=&  - f^{(1)}_{\theta \theta} \\
f^{(2)}_{\phi \phi}     &=&  3 \sin^2 \theta - 1  .
\end{eqnarray}

We present results for a gravitational wave crossing two fixed mesh refinement
boundaries into regions with successively coarser resolution.  We start with a
wave packet composed of a linear combination of one initially ingoing and one
outgoing
wave, each having amplitude $A = 10^{-6}$ and width
$\omega = 1$.  This packet is centered on the origin in a fine grid region of 
resolution $h = 0.0416667$.  The successively coarser regions have resolutions
$2h$ and $4h$, with the first refinement boundary at 
$r = \sqrt{x^2+y^2+z^2}=4.5$ and the second at $r = 9.0$.  To complete the initial
data we take $K_{ij} = 0$ so that $K = 0$ and $\tilde A_{ij} = 0$.  Octant
symmetry is imposed by using mirror-symmetry boundary conditions along
$x = 0$, $y = 0$, and $z = 0$.  The computational domain covers the regions
$0 \le x \le 12$ and similarly for $y$ and $z$.

As the evolution proceeds, the outgoing waves travel directly toward the outer
boundary of the grid.  The initially ingoing waves first travel toward the
origin, then reflect and move outward.  As the overall signal propagates outward,
it leaves flat spacetime behind.  

\begin{figure}[!ht]
\centerline{\epsfxsize=14cm\epsfysize=21cm\epsffile{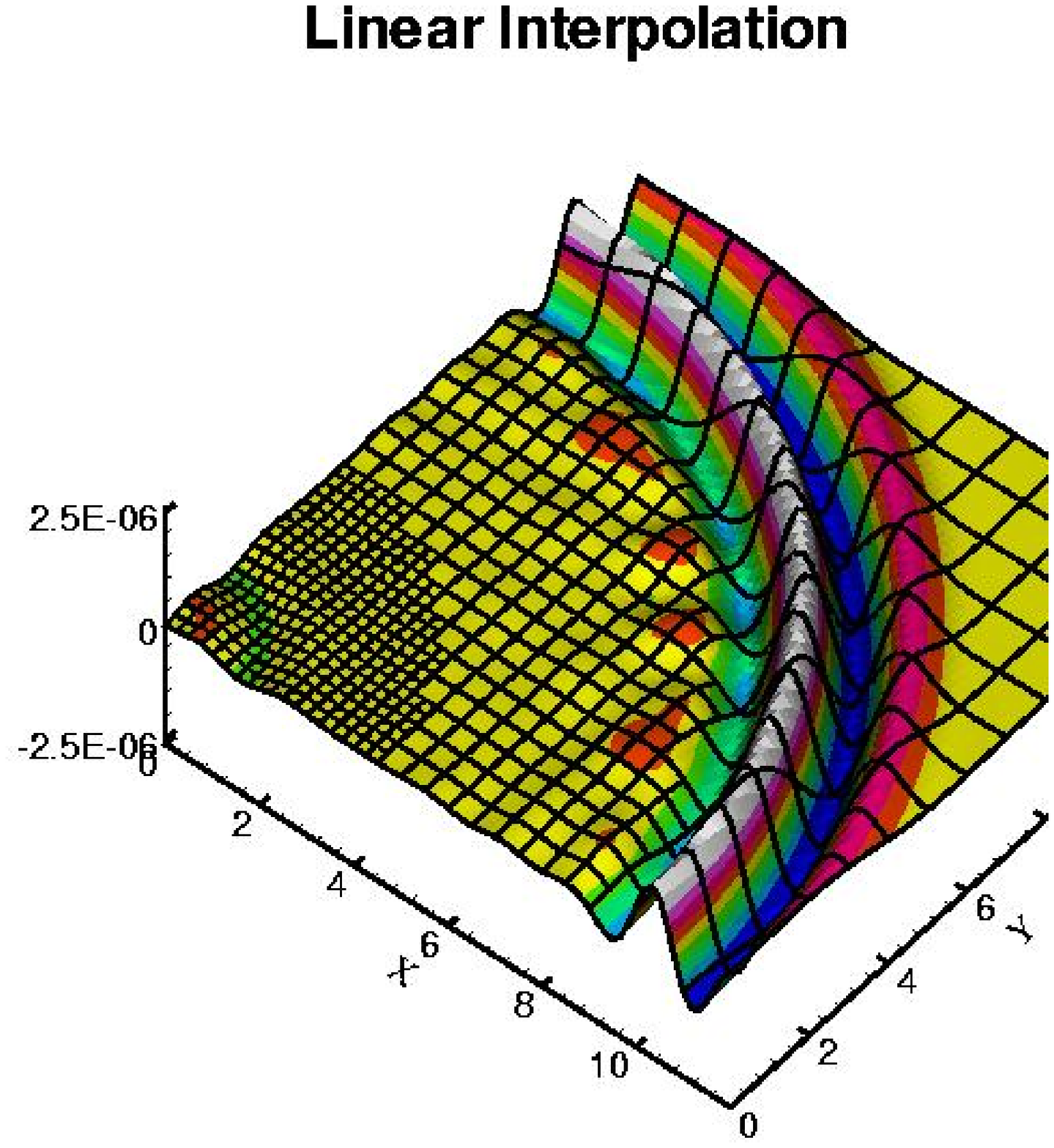}}
\caption{Evolution of gravitational waves in the 3--D Einstein equations using 
linear GCF.  
$g_{zz}-1$ is plotted 1n the $z=0$ plane.
Three levels of resolution (h,2h,4h)  are used, with $h = 0.0416667$.
Each of the grid blocks shown has $6 \times 6 \times 6$ zones.
}
\label{3d_2dcut_lin}
\end{figure}

\begin{figure}[!ht]
\centerline{\epsfxsize=14cm\epsfysize=21cm\epsffile{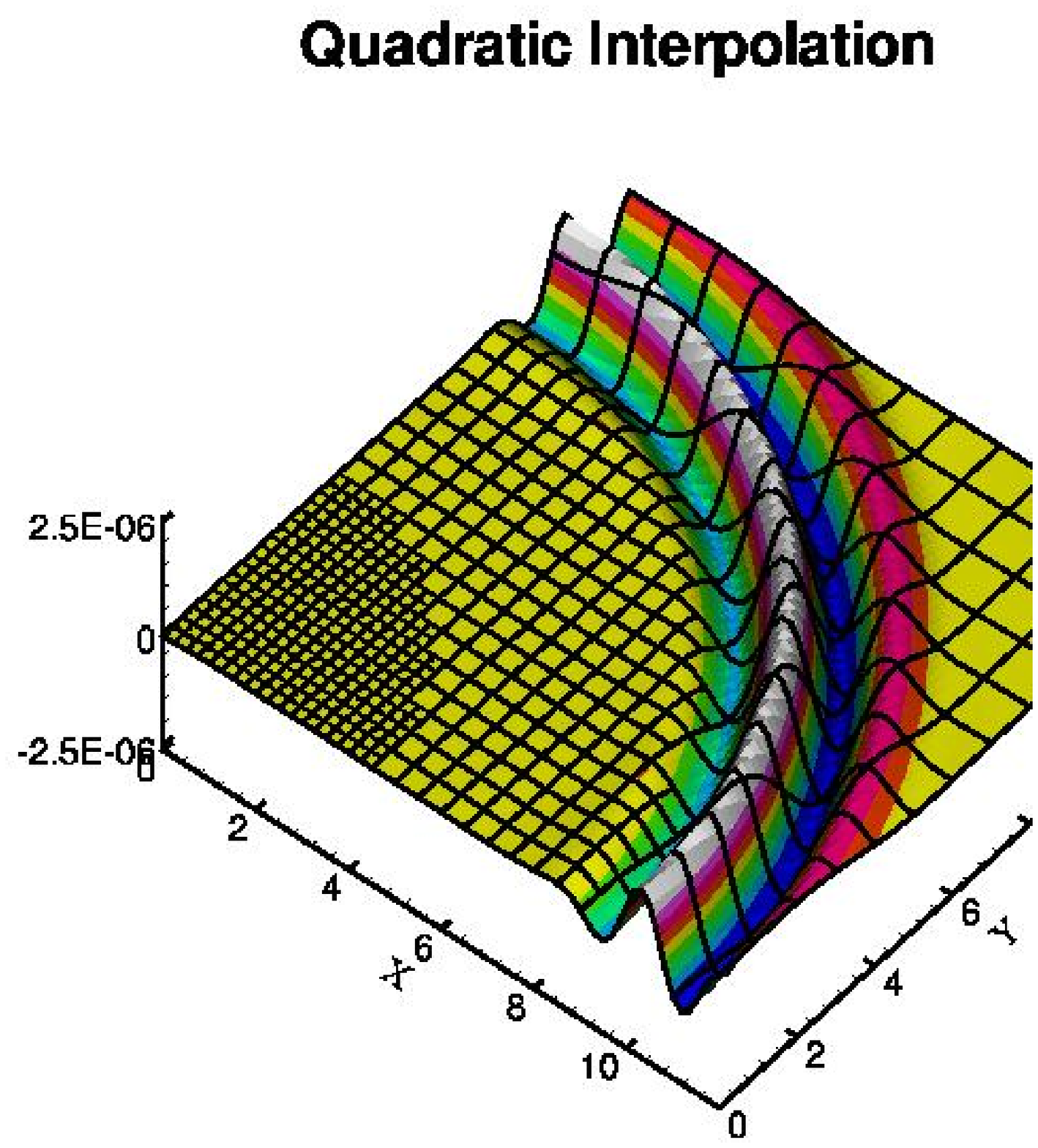}}
\caption{Same as Fig.~\ref{3d_2dcut_lin} except that quadratic
GCF is used.
}
\label{3d_2dcut_quad}
\end{figure}

\begin{figure}[!htb]
\includegraphics[scale = 0.75]{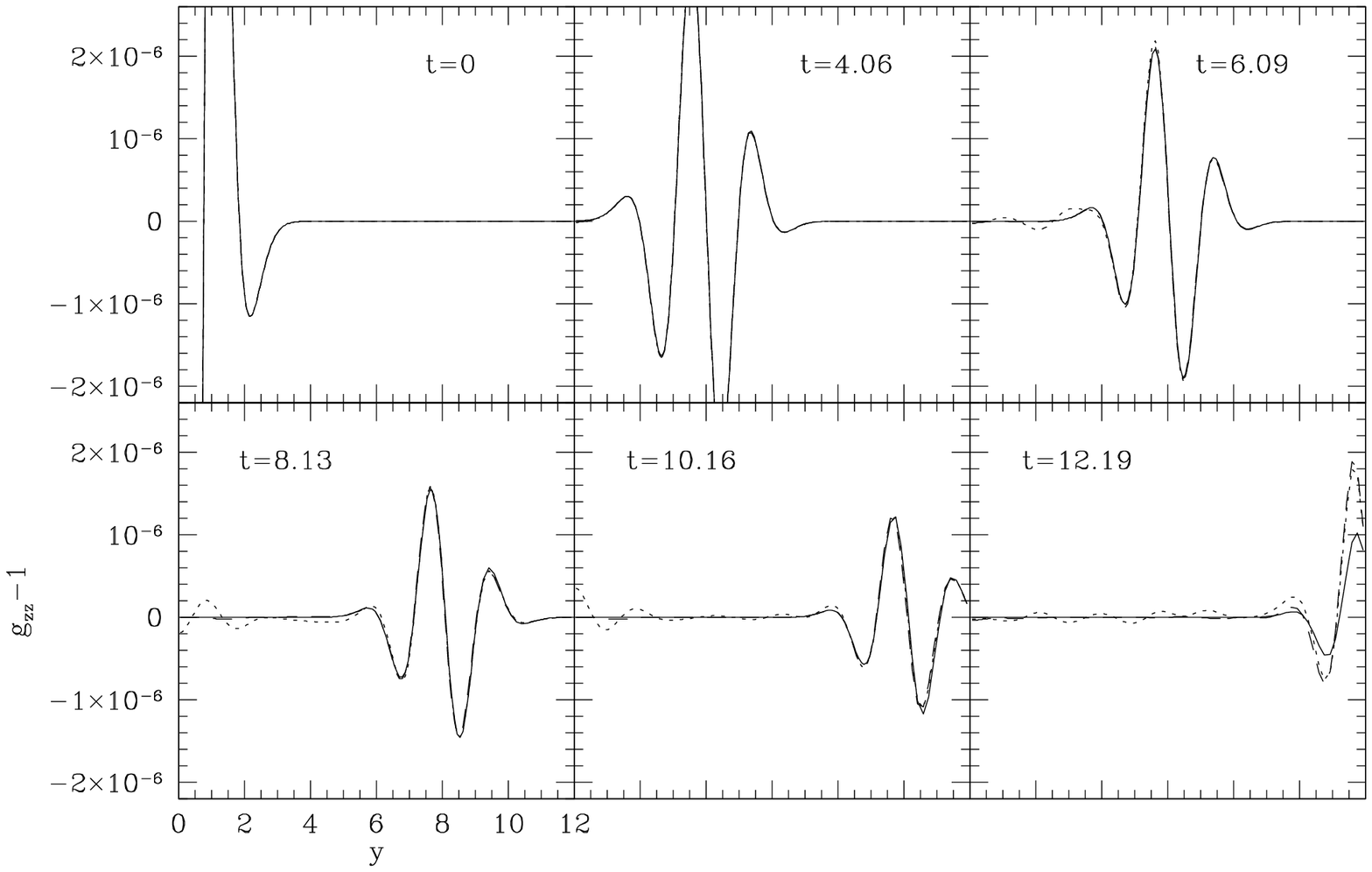}
\caption{The evolution of $g_{zz}-1$ is shown along the $y$-axis for 
the 3--D Einstein equations.
The analytic solution (solid line) and numerical solutions using linear
(dotted line) and quadratic (dashed line) GCF are shown.  The resolution
levels are given by (h,2h,4h) with $h = 0.0416667$.}
\label{3d_1dcut}
\end{figure}

Figure~\ref{3d_2dcut_lin} shows the evolution of these waves when linear GCF
is used.  The function $g_{zz} - 1$ is plotted as
a function of $x$ and $y$ in the $z=0$ plane at 4 successive times.
Note the presence of spurious reflected signals as the waves pass through the
fixed mesh boundaries.  These problems are greatly reduced when quadratic 
GCF is used, as shown in Fig.~\ref{3d_2dcut_quad}.
A comparison of runs with linear (dotted line) and quadratic (dashed line) 
GCF and the analytic solution (solid line) is shown in Fig.~\ref{3d_1dcut}.
The reflected waves are essentially eliminated by the use of quadratic GCF.

\begin{figure}[!htb]
\includegraphics[scale=0.75]{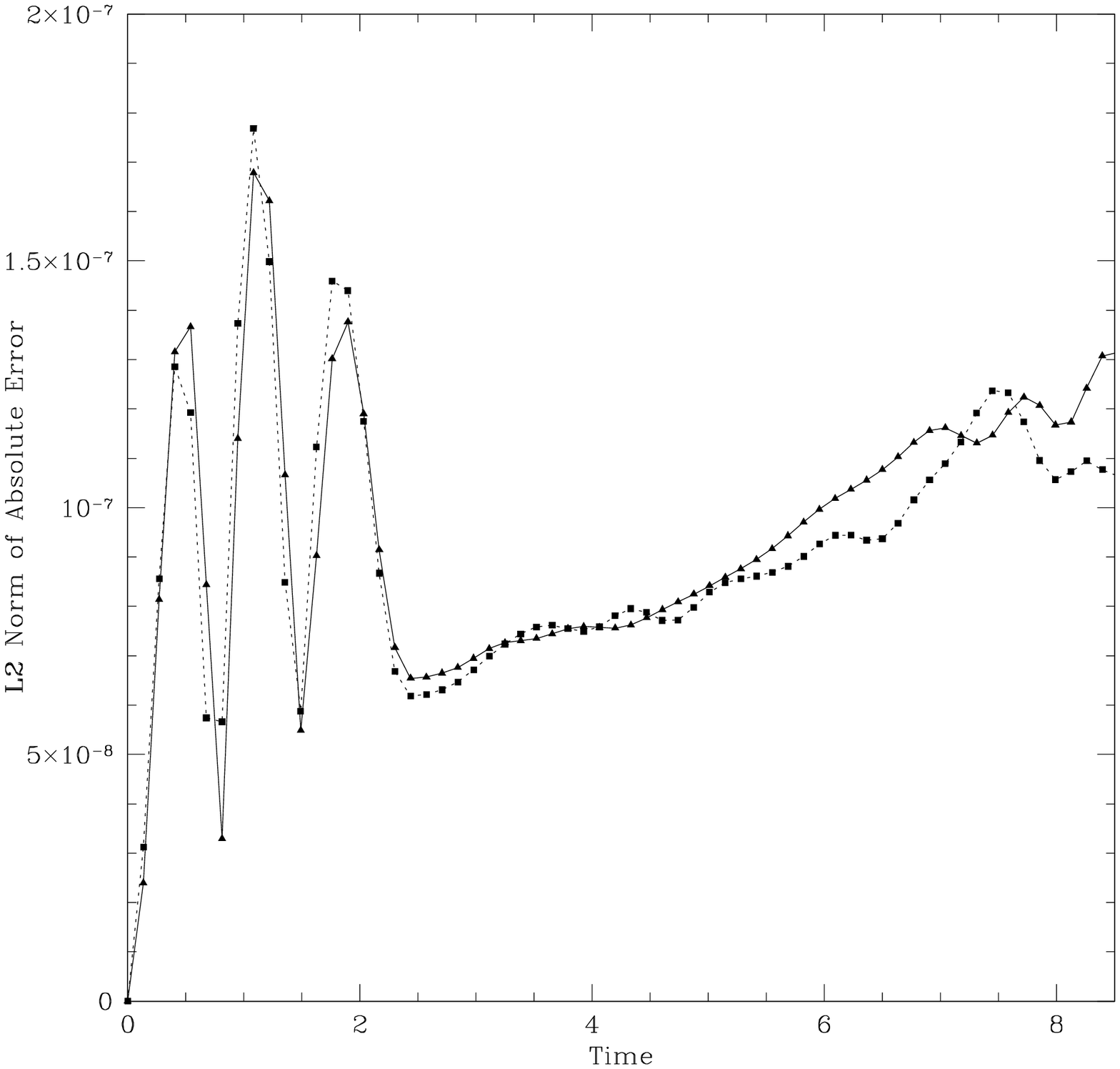}
\caption{The {\tt L2} norm of the absolute error $\epsilon$ 
is shown for two gravitational wave
runs differing in resolution by a factor of 2 throughout.  Both runs have
2 refinement boundaries and use quadratic GCF.  The solid triangles connected
by the solid line show $\epsilon$ for the run at higher overall resolution, and
the solid boxes connected by the dotted line show $4 \times \epsilon$ for the run
with lower overall resolution.}
\label{3d_convtest}
\end{figure}

Finally, Fig.~\ref{3d_convtest} demonstrates the second--order convergence
of the code by comparing the results of the run in Fig.~\ref{3d_2dcut_quad}
with a run that differs only by having the size of the grid zones a factor
of 2 larger throughout.  Both runs use quadratic GCF.  The L2 norm of the
absolute error $\epsilon$ is calculated over each simulation domain, and plotted
as a function of time.  The solid triangles connected by the solid line
show $\epsilon$ for the run in Fig.~\ref{3d_2dcut_quad}, and the filled
boxes connected by the dotted line show the errors for the lower resolution
run multiplied by 4.

\section{Summary}
\label{summary}

We have examined the propagation of waves across fixed mesh refinement
boundaries, starting with simplified linear and nonlinear model equations
in 1--D and 2--D, and progressing to the 3--D Einstein equations of general
relativity.  The numerical evolutions were carried out using centered spatial
differences and the explicit iterated Crank-Nicholson time update method, 
giving second--order accuracy.  Our results show that using linear GCF produces
spurious reflected waves as the signals cross refinement boundaries, and that
these are greatly suppressed by using quadratic GCF.  In particular, 
quadratic GCF preserves the second--order convergence of the numerical
evolutions.  Our numerical results are complemented by a detailed analytic
treatment of waves crosing refinement boundaries in 1--D in the Appendix.

While quadratic GCF is straightforward to describe and implement in 1--D,
the situation becomes more complicated in 2--D.  In particular, intermediate
values parallel to the mesh refinement interface must be calculated in the 
2--D case.  We have found that using linear interpolation to obtain these
intermediate values, combined with quadratic interpolation for the final
values, maintains the second--order convergence.  The procedure used for
quadratic GCF in 2--D generalizes in a straightforward manner to the
3--D case.

The techniques presented here appear to be robust in the sense that they continue to
produce excellent results as our test problems increase in complexity.  Quadratic
GCF successfully eliminates most of the spurious reflected waves in both linear and nonlinear
model equations in 1--D and 2--D.  The 3--D Einstein equations present a much 
larger and more complex system of equations.  In the test case presented here,
the evolution of a weak gravitational wave, quadratic GCF continues to perform
well, even as the signals cross two successive mesh refinement boundaries.
We fully expect that these techniques will also yield excellent results for
strong gravitational waves, which activate the nonlinear terms in the Einstein
equations.  Such evolutions require various technical differences in the 
gauge choices ($\alpha$ and $\beta^i$) as well as in the formulation of the
initial data.  We are currently working on such models, and will report on
them elsewhere.

\ack

It is a pleasure to thank John Baker, Phillip Colella, Kevin Olson, and
Steve Zalesak for helpful and stimulating discussions.
The work was supported in part by NSF grant PHY-0070892.


\appendix

\section{Appendix: Analysis of Numerical Wave Propagation in 1-D}
\label{analysis}

In this appendix, we present a more detailed analysis of the propagation
of linear waves in 1-D with the discretization described 
in Sec.~\ref{lin-1d}.  We begin by deriving some basic results for uniform 
grids (Sec.~\ref{analysis-uniform}) and follow with  a study of wave propagation 
across a fixed mesh refinement boundary (Sec.~\ref{analysis-FMR}). Here, we do 
not address the issue of instabilities that might arise due to coupling 
between fine and coarse meshes \cite{OandP}. However, as noted in Sec.~4, our numerical 
tests show no signs of instability. 

\subsection{Wave Propagation on a Uniform Mesh}
\label{analysis-uniform}

We will employ matrix notation to facilitate the analysis in this
Appendix.  First, we collect the field variables $\phi$, 
$\Pi$ into the column vector
\begin{equation}
	V = \left(\matrix{\phi \cr \Pi}\right) \ .
\end{equation}
Equations (\ref{eq:phi})--(\ref{eq:pi}) can now be written as
\begin{equation} 
	{\partial V \over \partial t} = 
	\left(\matrix{0 & 1 \cr {\partial^2/\partial x^2} & 0}\right) V \ .
\end{equation}
As usual $V^n_j$ will denote the vector of grid functions at
timestep $n$ and grid point $j$. 

The iterated Crank--Nicholson method described in Sec.~\ref{discretiz} is built
from successive applications of the basic operator 
\begin{equation}
	Q = \left( \matrix{0 & 1 \cr \partial^2 & 0} \right) \ ,
\label{Q}
\end{equation}
where $\partial^2 V^n_j \equiv (V^n_{j+1} - 2V^n_j + V^n_{j-1})/\Delta x^2$.
With two iterations, the update of the variables $V^n_j$ 
by one full timestep is accomplished by the operator
\begin{equation}
	M = I + \Delta t\, Q\left[ I + {\Delta t\over 2}Q\left( 
	I + {\Delta t\over 2}Q\right)\right] \ .
\label{M1}
\end{equation}
The stability, dissipation, and dispersion properties are obtained 
by considering discrete plane wave solutions, 
\begin{equation} 
	V^n_j = W e^{i\omega n \Delta t} e^{- i k j \Delta x} \ ,
\label{V-ansatz}
\end{equation}
where $W$ is a constant vector (independent of $n$ and $j$). Inserting 
this ansatz into the update equation $V^{n+1}_j = M V^n_j$, we find
\begin{equation}
	e^{i\omega \Delta t} W = 
	\left( \matrix{1 - 2\Lambda^2 & \Delta t(1 - \Lambda^2) \cr
	-4\Lambda^2(1 - \Lambda^2)/\Delta t & 1 - 2\Lambda^2 }\right) W \ ,
\label{eW}
\end{equation}
where 
\begin{equation}
	\Lambda \equiv {\Delta t\over \Delta x} \sin(k\Delta x/2) \ .
\label{lambda}
\end{equation}
Thus, $W$ is an eigenvector with eigenvalue 
$e^{i\omega \Delta t}$ for the matrix that appears in Eq.~(\ref{eW}). 
The eigenvalues are obtained in the usual way with the result
$ e^{i\omega \Delta t} = 1 - 2\Lambda^2 \pm 2i\Lambda(1 - \Lambda^2)$.
This is the dispersion relation giving the complex frequency $\omega$ as a 
function of wave number $k$. We can, without loss of generality, consider 
only plane wave solutions (\ref{V-ansatz}) with positive frequency $\xi > 0$, where 
$\xi = \Re(\omega)$ is the real part of $\omega$. Then the $\pm$ sign in the 
dispersion relation must be set equal to the sign of the wave number $k$. 
The dispersion relation then becomes 
\begin{equation}
	e^{i\omega \Delta t} = 1 - 2\Lambda^2 + 2i|\Lambda|(1 - \Lambda^2) 
\label{dispersion-2}
\end{equation}
and $\xi = \Re(\omega)$ is positive. 
The eigenvectors $W$ corresponding to these eigenvalues are 
straightforward to compute. Choosing the first component of $W$ to be unity, 
we find
\begin{equation}
	W = \left(\matrix{1 \cr 2i|\Lambda|/\Delta t }\right) \ .
\label{W-vals}
\end{equation}
We note for later reference that $W e^{-ikj\Delta x}$ is an eigenvector of the 
basic operator $Q$ with eigenvalue $2i|\Lambda|/\Delta t$.

The finite difference scheme is unstable if the magnitude of the amplification 
factor, $|e^{i\omega\Delta t}|$, is greater than unity. From Eq.~(\ref{dispersion-2}) 
we find that $|e^{i\omega\Delta t}|^2 \leq 1$ implies $\Lambda^2 \leq 1$. This 
inequality will be satisfied for all wave numbers $k$ only if $\Delta t \leq \Delta x$. 
This is the Courant limitation on the timestep for the wave equation 
(\ref{eq:phi})--(\ref{eq:pi}) discretized with 
twice--iterated Crank--Nicholson. 

The phase velocity is found from the real part of the frequency $\xi = \Re(\omega)$. From 
the dispersion relation (\ref{dispersion-2}), we find 
\begin{equation}
	\xi\Delta t = \arcsin\left({2|\Lambda|(1 - \Lambda^2) \over 
	\sqrt{1 - 4\Lambda^4 (1 - \Lambda^2)}}\right) \ .
\end{equation}
The phase velocity is then
\begin{equation}
	c(\lambda) = {\xi\over k} = {\xi\Delta t \over 2\pi\alpha} 
	{\lambda \over \Delta x} \ ,
\label{phase-veloc}
\end{equation}
where $\alpha \equiv \Delta t/\Delta x$ is the Courant factor and $\lambda = 2\pi/k$ 
is the wavelength. The dissipation is found from the 
imaginary part of the frequency, $\eta = \Im(\omega)$. Since the wave amplitude 
varies like $\phi \sim e^{-\eta n \Delta t}$, we see that 
the amplitude drops by a factor 
\begin{equation}
    e^{-\eta \Delta t} = |e^{i\omega \Delta t}|
   = \sqrt{1 - 4\Lambda^4 (1 - \Lambda^2)} \ .
\label{dissip}
\end{equation}
for each timestep.

\subsection{Wave Propagation with FMR}
\label{analysis-FMR}

Now consider a two--level refined mesh, with fine grid $\Delta x_f$ on the 
left and coarse grid $\Delta x_c$ on the right.  We will assume that the 
refinement jumps by a factor of $2$, that is, $\Delta x_c = 2\Delta x_f$. 
The mesh will be labeled as shown in Fig.~\ref{fig:guard_cell}. 
Thus, $V^n_{-1/2}$, $V^n_{-3/2}$, {\it etc.} 
are the fine grid functions and $V^n_{1/2}$, $V^n_{3/2}$, {\it etc.} are the 
coarse grid functions.

As a first step towards analyzing the wave reflection and transmission at the 
interface, we relate the wave numbers in the coarse and fine grid regions. 
Consider a monochromatic solution that varies like $\phi \sim e^{i\xi n\Delta t}$ 
across the entire mesh. Specifically, we assume that the coarse and 
fine grid frequencies are the same, $\xi_c = \xi_f$, and that the 
coarse and fine grid time steps are the same, $\Delta t_c = \Delta t_f$. 
From the dispersion relation, Eq.~(\ref{dispersion-2}),
we can compute $\tan(\xi\Delta t)$ in the 
coarse and fine grid regions and equate the results:
\begin{equation}
	{2|\Lambda_c|(1 - \Lambda_c^2) \over 1 - 2\Lambda_c^2} 
	= {2|\Lambda_f|(1 - \Lambda_f^2) \over 1 - 2\Lambda_f^2} \ .
\label{equate}
\end{equation}
Here, $\Lambda_c = (\Delta t/\Delta x_c)\sin(k_c\Delta x_c/2)$ and similarly 
for $\Lambda_f$. This relation has the form $f(|\Lambda_c|) = f(|\Lambda_f|)$ 
where $|\Lambda_c|$ and $|\Lambda_f|$ vary between $0$ and $1$. It is easy to 
show that 
the function $f(|\Lambda|)$ is monotonic and therefore invertible. It follows that 
the only solution of Eq.~(\ref{equate}) is 
\begin{equation}
	|\Lambda_c| = |\Lambda_f| \ .
\label{mods-cf}
\end{equation}
This equation shows that the coarse and fine grid wave numbers $k_c$ 
and $k_f$ are related by
\begin{equation}
	k_c = \pm {2\over \Delta x_c} \arcsin\left[ {\Delta x_c\over \Delta x_f}
	\sin(k_f\Delta x_f/2) \right] \ .
\label{kc}
\end{equation}
The two cases corresponding to the $\pm$ sign indicate that the wave propagation direction 
on the coarse and fine sides of the interface need not match. Thus, 
we can have a right moving wave in the coarse grid region connected to both 
right moving and left moving waves in the fine grid region. 

From the result~(\ref{mods-cf}) we see that the rate of dissipation~(\ref{dissip}) 
of a wave, governed by $\eta = \Im(\omega)$, 
is the same in the coarse and fine grid regions. We also see that the 
relative phase between the two components $\phi$ and $\Pi$ of the wave,
Eq.~(\ref{W-vals}), is the same in coarse and fine regions.
The phase velocity (\ref{phase-veloc}), and 
hence the amount of dispersion, differ in the coarse and fine grid regions, since the 
wave numbers $k_c$ and $k_f$ are not equal. 

According to Eq.~(A.15) $k_c$ is real only for $k_f \Delta x_f \leq \pi/3$, 
that is, for $\lambda_f/\Delta x_f \geq 6$. If $\lambda_f/\Delta x_f < 6$, then 
$k_c$ is complex and the plane wave solution (A.5) will contain a spatial dependence 
in the coarse grid region that is either exponentially damped or grows exponentially. 
Note that, although $k_c$ might be complex, $\Lambda_c$ 
is real (assuming $k_f$ is real) and equal to $\pm \Lambda_f$. It follows that, 
whether $k_c$ is real or complex, the 
Courant stability condition $|e^{i\omega\Delta t}|^2 \leq 1$ is satisfied in the 
coarse grid region if it is satisfied in the fine grid region. 

For the remainder of this appendix we will focus on the case of practical interest, 
where $\lambda_f/\Delta x_f \geq 6$ and $k_c$ is real. The plots in Figures 
5, 6, and 7 have been restricted to $\lambda_f/\Delta x_f \geq 10$ for 
clarity of presentation. Each of the curves in those plots reaches a finite value 
at $\lambda_f/\Delta x_f = 6$.

\subsubsection{Matching solutions}
\label{match-f2c}

At this point we have shown that waves of frequency $\xi$ have wave number $k_c$ 
on a coarse grid, wave number $k_f$ on a fine grid, and that these values are related 
as in Eq.~(\ref{kc}). We will now construct a solution with frequency $\xi$ that 
spans the entire non-uniform grid. To begin, consider the vector 
\begin{equation}
	V_j = \cases{ W\left( e^{-ik_f j\Delta x_f} 
	+ {\bf R}e^{ik_f j \Delta x_f} \right) \ ,
	& $j < 0$ \ ;\cr
	W\left( {\bf T} e^{-ik_c j\Delta x_c} \right) \ , & $j > 0$ \ . \cr}
\label{ansatz-2}
\end{equation}
We will show that for an appropriate 
choice of the coefficients ${\bf R}$ and ${\bf T}$ the vector $V_j$ is an eigenvector of the 
basic operator $Q$ with eigenvalue $2i|\Lambda|/\Delta t$. 
For points away from the interface, namely, the points $j \le -3/2$ and $j \ge 3/2$, 
this conclusion follows from our earlier observation that on a uniform grid 
$W e^{\pm ikj\Delta x}$ is an eigenvector of $Q$ with eigenvalue $2i|\Lambda|/\Delta t$.
The same argument cannot be applied to the points $1/2$ and $-1/2$ surrounding the interface 
because the stencil for the discrete derivative operator $\partial^2$ appearing in $Q$ 
extends across the interface. Thus, when 
computing $\partial^2 V_j$ for $j = \pm 1/2$, we must use guard cell 
information. 

In the main text we discussed  various choices for guard cell filling, such as the 
Paramesh linear GCF of Eqs.~(\ref{fG-lin})--(\ref{fg-lin}) and the quadratic GCF of 
Eqs.~(\ref{fg-quad})--(\ref{fG-quad}). 
For the purpose of presenting the analysis, we will focus 
instead on the direct linear GCF
of Eqs.~(\ref{fG-direct})--(\ref{fg-direct}). In the present notation, 
these relations are
\begin{equation}
	V^n_G  =  {1\over 2}  (V^n_{-3/2} +  V^n_{-1/2}), \qquad 
	V^n_g  =  {1\over 3}  (V^n_{-1/2} + 2 V^n_{1/2}) 
\label{dgcf}
\end{equation}
Now, for grid points 
that are not adjacent to the interface, the operator $\partial^2$ takes the usual form, 
\begin{equation}
	\partial^2 V^n_j = \left\{ \matrix{ 
	(V^n_{j+1} - 2V^n_j + V^n_{j-1})/\Delta x_f^2 \ ,& j \le -3/2 \ , \cr
	(V^n_{j+1} - 2V^n_j + V^n_{j-1})/\Delta x_c^2 \ ,& j \ge 3/2 \ .} \right.
\end{equation}
But for grid points adjacent to the interface, $\partial^2$ must use guard cell 
values given by (\ref{dgcf}). Consequently, we find 
\begin{eqnarray}
	\partial^2 V^n_{-1/2}  & = &
  	(V^n_g - 2V^n_{-1/2} + V^n_{-3/2})/\Delta x_f^2 \nonumber \\
	& = & (2V^n_{1/2} - 5V^n_{-1/2} + 3V^n_{-3/2})/(3\Delta x_f^2) 
\label{d2-min-lin}
\end{eqnarray}
and
\begin{eqnarray}
	\partial^2 V^n_{1/2} & = &
	(V^n_{3/2} - 2 V^n_{1/2} + V^n_G)/\Delta x_c^2 \nonumber \\
	& = & (2V^n_{3/2} - 4 V^n_{1/2} + V^n_{-1/2} + V^n_{-3/2})/(2\Delta x_c^2) \ .
\label{d2-p-lin}
\end{eqnarray}
for $\partial^2$ acting at grid points $j = \pm 1/2$. 

We now impose the requirement that the vector $V_j$ of Eq.~(\ref{ansatz-2}) is 
an eigenvector of $Q$ with eigenvalue $2i|\Lambda|/\Delta t$ at the points $j = \pm 1/2$ 
adjacent to the interface. Using the discretization (\ref{d2-min-lin}) the 
relation $QV_{-1/2} = (2i|\Lambda|/\Delta t)V_{-1/2}$ yields 
\begin{equation}
	{1\over 3}(2\phi_{1/2} - 5\phi_{-1/2} + 3\phi_{-3/2}) = 
	\left({2i|\Lambda| \over \alpha_f}\right)^2 \phi_{-1/2} \ .
\label{rel-40}
\end{equation}
Here, $\phi_j$ is the first component of the ansatz vector $V_j$. 
Similarly, with the discrete operator (\ref{d2-p-lin}), we find that 
$QV_{1/2} = (2i|\Lambda|/\Delta t)V_{1/2}$ implies 
\begin{equation}
	{1\over 2}(2\phi_{3/2} - 4\phi_{1/2} + \phi_{-1/2} + \phi_{-3/2}) = 
	\left({2i|\Lambda| \over \alpha_c}\right)^2 \phi_{1/2} \ .
\label{rel-41}
\end{equation}
These two equations can be solved for the two coefficients 
${\bf R}$ and ${\bf T}$. The result is 
\begin{eqnarray}
	{\bf R} & = & {3 E_c^2 E_f^2 - E_f^4 - E_c^2 E_f^4 - E_f^6 \over 
	1 + E_f^2 + E_c^2 E_f^2 - 3E_c^2 E_f^4} \ ,\nonumber \\
	{\bf T} & = & {E_c(3 +2E_f^2 - 2E_f^6 - 3E_f^8)/(2E_f) \over
	  1 +E_f^2 + E_c^2 E_f^2 - 3E_c^2 E_f^4} \ ,
\label{direct-RT}
\end{eqnarray}
where the shorthand notation 
\begin{equation}
	E_c \equiv e^{ik_c\Delta x_c/2} \ ,\qquad 
 	E_f \equiv e^{ik_f\Delta x_f/2}
\end{equation}
has been used. 

At this point we have succeeded in showing that the vector $V_j$ of 
Eq.~(\ref{ansatz-2}), with ${\bf R}$ and ${\bf T}$ chosen as in 
Eqs.~(\ref{direct-RT}), is an eigenvector of $Q$ 
on the non--uniform grid. The corresponding eigenvalue is 
$2i|\Lambda|/\Delta t$. A short calculation shows that $V_j$ is also an 
eigenvector for $M$, Eq.~(\ref{M1}),  with
eigenvalue $1 - 2\Lambda^2 + 2i|\Lambda|(1 - \Lambda^2)$. Since $M$ evolves the 
discrete system by one time step, we see that 
\begin{equation}
	V_j^n = \cases{ We^{i\omega n\Delta t}\left( e^{-ik_f j\Delta x_f} 
	+ {\bf R}e^{ik_f j \Delta x_f} \right) \ ,
	& $j < 0$ \ ;\cr
	We^{i\omega n\Delta t}\left( {\bf T} e^{-ik_c j\Delta x_c} \right) 
	\ , & $j > 0$ \ .\cr}
\label{fmr-solution}
\end{equation}
is a solution of the finite difference equations $V_j^{n+1} = MV_j^n$ across the 
entire grid. Here, the complex frequency $\omega$ is given by the dispersion 
relation~(\ref{dispersion-2}). The solution (\ref{fmr-solution}) represents a 
wave that travels from the fine grid region to the coarse grid region. At the 
interface it splits into 
reflected and transmitted pieces. The coefficient ${\bf R}$ is the reflection 
coefficient, and ${\bf T}$ is the transmission coefficient. 

\begin{figure}[!htb]
\includegraphics{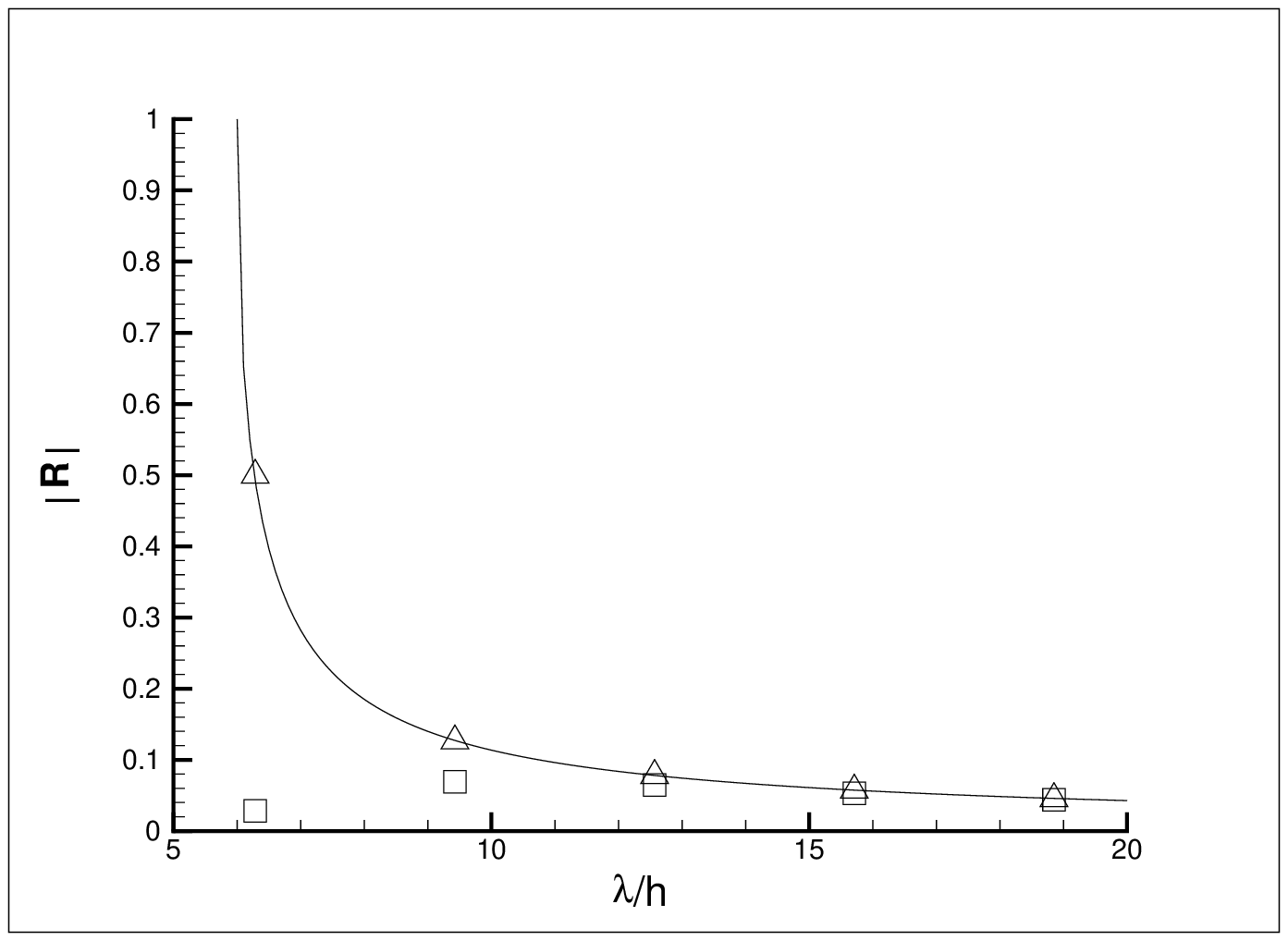}
\caption{Absolute value of the reflection coefficient for direct linear 
guard cell filling. The raw numerical data are shown as boxes, while the triangles show 
the data corrected for dissipation.}
\label{refl-test}
\end{figure}

The results of this analysis can be checked numerically. For example, in 
Fig.~\ref{refl-test} the solid lines show the magnitude of 
the reflection coefficient calculated from Eq.~(\ref{direct-RT}). 
The squares and triangles display the results of a numerical test, in which 
we modeled the propagation of a sine wave contained in a
broad Gaussian envelope, 
\begin{equation}
	\phi(x,t) = A e^{-(x-x_0-t)^2/w^2} \sin{k(x-t)} \ .
\end{equation}
The initial data were obtained by discretizing $\phi$ and $d\phi/dt$ at $t=0$. The constant 
$x_0$ was chosen so that initially the wave packet was situated in the fine grid region, 
far from the interface. We set the Gaussian half--width to $w = 63\Delta x$, 
significantly larger than the longest wavelength shown in the figures. The magnitude of 
${\bf R}$ was obtained by evolving the wave packet past its interaction 
with the interface, typically about 1000 time steps, and then extracting from the numerical 
data the largest value of $|\phi|$ in the reflected pulse. 
The Courant factor for these numerical runs was $\Delta t/\Delta x_f = 0.4$.

The squares show raw numerical data. The deviation of these data from the analytical 
curves is due to dissipation and dispersion of the wave pulse. Dissipation can be 
accounted for rather easily, using the analytical result given in
Eq.~(\ref{dissip}). The triangles show the 
numerical data with correction for dissipation. The numerical and analytical 
results for the reflection coefficient are in excellent agreement, as seen in 
Fig.~(\ref{refl-test}).

The above analysis can be repeated for other choices of GCF. In particular, for 
the quadratic GCF of Eqs.~(\ref{fg-quad})--(\ref{fG-quad}), the reflection and 
transmission coefficients are 
\begin{eqnarray}
        {\bf R} & = & {E_f^2(-1 + 6E_f^2 + 3E_f^4 + E_c^2(-15 + 10E_f^2 - 3E_f^4)) \over
        3 +6E_f^2 - E_f^4 - E_c^2(3 - 10E_f^2 + 15E_f^4)} \ ,\nonumber\\
        {\bf T} & = & {2E_c(-3 - 2E_f^2 + 2E_f^6 + 3E_f^8)/E_f \over
        3 + 6E_f^2 - E_f^4 - E_c^2(3 - 10E_f^2 + 15E_f^4) }\ ,
\label{quad-RT}
\end{eqnarray}
with $E_c$ and $E_f$ defined as before. 
Figures (\ref{R_compare})--(\ref{Tphase_compare}) compare the 
reflection and transmission coefficients for linear,
direct linear, and quadratic GCF. 
From Fig.~\ref{R_compare} we see that quadratic GCF produces a much 
smaller reflected wave than either linear or 
direct linear GCF. For example, at $20$ 
fine grid points per wavelength, linear (direct linear) GCF
produces a reflected wave with amplitude about $5.6\%$ ($4.3\%$) that of the 
incident wave. With quadratic GCF the reflected wave amplitude is only 
about $0.31\%$ that of the incident wave. For the transmitted wave, 
Fig.~\ref{Tphase_compare} shows that the phase error at $20$ fine grid points per wavelength 
is smaller in magnitude by more than a factor of $10$ with 
quadratic GCF  compared to linear or direct linear GCF. Figure \ref{T_compare} shows, 
perhaps surprisingly, that direct
linear GCF does the best  job of keeping the magnitude of ${\bf T}$ 
close to $1$ at wavelengths less than about $28\Delta x_f$. For longer wavelengths, 
$\lambda > 28\Delta x_f$, quadratic GCF is best. Note, however, that with 
any of these choices of GCF, the deviation of $|{\bf T}|$ away 
from unity is relatively small. Over the entire range shown in the graphs, 
$\lambda > 10\Delta x_f$, the maximum error for quadratic GCF 
is less than $1\%$. For the reflection coefficient, 
on the other hand, the maximum error for direct linear GCF is about $11\%$. 
For most situations the problem of 
spurious reflections at an interface will be more severe than the problem of 
inaccurate wave
transmission through the interface. 

\subsubsection{Discussion of guard cell filling}
\label{discuss-gcf}

The results thus far indicate that quadratic GCF is generally 
better than either linear or direct linear 
GCF at keeping the reflection coefficient small and the 
transmission coefficient close to unity. 
The question naturally arises: can one do better than quadratic GCF? We will restrict our 
attention to rules for GCF that use a three--point stencil. That is, the fine 
and coarse grid guard cell values $V_g^n$ and $V_G^n$ are obtained from linear combinations 
of three grid points, 
\begin{equation}
	V_G^n =  c_1 V^n_{1/2} + c_2 V^n_{-1/2} + c_3 V^n_{-3/2} \ ,
\label{VG}
\end{equation}
\begin{equation}
	V_g^n = f_1 V^n_{1/2} + f_2 V^n_{-1/2} + f_3 V^n_{-3/2} \ ,
\label{Vg}
\end{equation}
where $c_1$, $f_1$, {\it etc.}~are constants. 

The analysis of Sec.~\ref{match-f2c}  
can be repeated with the guard cells defined as above. 
The resulting reflection and transmission coefficients are functions of the constants 
$c_1$, $f_1$, {\it etc}. Although there are certain combinations of the constants that 
outperform quadratic GCF at low resolution, quadratic GCF 
is unique in the following sense. If we consider the high resolution limit, in which 
$k_f\Delta x_f$ is small, quadratic GCF yields
\begin{equation}
	{\bf R} =  {3\over 32}i(k_f\Delta x_f)^3 + {\bf O}(k_f\Delta x_f)^4 \ ,
\label{R-QGCF}
\end{equation}
\begin{equation}
	{\bf T} = 1 - {3\over 32}i(k_f\Delta x_f)^3 + {\bf O}(k_f\Delta x_f)^4 \ .
\label{T-QGCF}
\end{equation}
The magnitudes of the reflection and transmission coefficients behave like 
$|{\bf R}| = {\bf O}(k_f\Delta x_f)^3$ and $|{\bf T}| = 1 + {\bf O}(k_f\Delta x_f)^4$. 
For any other choice of constants $c_1$, $f_1$, {\it etc.}, the reflection and transmission 
coefficients approach $0$ and $1$, respectively, more slowly (if at all) than for quadratic
GCF. For example, for direct linear GCF, we have 
\begin{eqnarray}
	{\bf R} & = & {1\over 8}i(k_f\Delta x_f) + {\bf O}(k_f\Delta x_f)^2 \ ,\\
	{\bf T} & = & 1 + {1\over 8}i(k_f\Delta x_f) + {\bf O}(k_f\Delta x_f)^2 \ .
\end{eqnarray}
Thus, for high resolution, quadratic GCF is the best possible choice given 
the three--point stencil~(\ref{VG})--~(\ref{Vg}). Because ${\bf R}$ and ${\bf T}$ 
approach $0$ and $1$ rapidly, as the third power of $k_f\Delta x_f$, quadratic 
GCF performs well at all resolutions.



\begin{thebibliography}{99}

\bibitem{LIGO}B. Barish, First Generation Interferometers, in: J. Centrella, ed.,
 {\em Astropohysical Sources
for Ground-Based Gravitational Wave Detectors\/} (AIP, Melville, NW, 2001), 3

\bibitem{LIGO2}P. Fritschel, The Second Generation LIGO Interferometers,
in: J. Centrella, ed.,
 {\em Astropohysical Sources
for Ground-Based Gravitational Wave Detectors\/} (AIP, Melville, NW, 2001), 15

\bibitem{LISA}P. Bender, {\em et al.}, {\em LISA, Pre-Phase A Report}, 2nd. edition (1998),
unpublished (available online at: http://lisa.jpl.nasa.gov/documents/ppa2-09.pdf)

\bibitem{MTW}C. Misner, K. Thorne, and J. Wheeler, {\em Gravitation} (W. H. 
Freeman, New York, 1973)

\bibitem{SN}M. Shibata and T. Nakamura, Evolution of three-dimensional gravitational waves: Harmonic slicing case, {\em Phys. Rev.} {\bf D52}, 5428 (1995).

\bibitem{BS}T. Baumgarte and S. Shapiro, Numerical integration of Einstein's field equations, {\em Phys. Rev.} {\bf D59}, 024007 (1998).

\bibitem{choptuik93}M. Choptuik, Universality and scaling in gravitational collapse of a massless scalar field, {\em Phys. Rev. Lett.} {\bf 70}, 9 (1993).

\bibitem{slicebh}B. Br\"{u}gmann, Adaptive mesh and geodesically sliced Schwarzschild spacetime in 3+1 dimensions, {\em  Phys. Rev.} {\bf  D54}, 7361 (1996).


\bibitem{brugbbh}B. Br\"{u}gmann, Binary Black Hole Mergers in 3d Numerical Relativity, {\em Int. J. Mod. Phys.} {\bf D8}, 85 (1999).

\bibitem{PSW}P. Papadopoulos, E. Seidel, and L. Wild, Adaptive computation 
of gravitational waves from black hole interactions, {\em Phys. Rev.} {\bf D58}, 084002 (1998).

\bibitem{Newetal}
K. New, D. Choi, J. Centrella, P. MacNeice, M. Huq, and K. Olson,
Three-dimensional adaptive evolution of gravitational waves in numerical relativity,
{\em Phys. Rev.\/} {\bf D62}, 084039 (2000).

\bibitem{hern-thesis}S. Hern, {\em Numerical Relativity and Inhomogeneous Cosmologies},
(Ph.D. Thesis, Dept. Applied Maths. and Theoretical Physics, 
Cambridge University, Cambridge, UK, 2000)
(http://arxiv.org/abs/gr-qc/0001070).

\bibitem{BergerOliger}
M.J.~Berger and J.~Oliger, Adaptive mesh refinement for hyperbolic partial differential 
equations, {\em J. Computat. Phys.} {\bf 53}, 484 (1984).

\bibitem{BergerColella}
M.J.~Berger and P.~Colella, Local adaptive mesh refinement for shock hydrodynamics, 
{\em J. Comput. Phys.} {\bf 82}, 64 (1989).

\bibitem{berger87}M. J. Berger, On conservation at grid interfaces, 
{\em SIAM J. Numer. Anal.\/} {\bf 24}, 967 (1987).

\bibitem{Chesshire90} G. Chesshire and W. D. Henshaw, {\em J. Comput. Phys.} {\bf 90}, 1-64 (1990).

\bibitem{Henshaw03} W. D. Henshaw and D. W. Schwendeman, submitted to {\em J. Comput. Phys.} (2003).


\bibitem{numrec}W. Press, S. Teukolsky, W. Vetterling, and B. Flannery, 
{\em Numerical Recipies ($2^{\rm nd}$ edition)} (Cambridge University Press,
New York, 1994).

\bibitem{ICN}S. Teukolsky, On the stability of the iterated Crank-Nicholson method in
 numerical relativity, {\em Phys. Rev.} {\bf D61}, 087501 (2000).

\bibitem{paramesh}
P. MacNeice, K. Olson, C. Mobarry, R. de Fainchtein, and C. Packer,
PARAMESH: A parallel adaptive mesh refinement community toolkit,
{\em Computer Physics Comm.\/} {\bf 126}, 330 (2000). 

\bibitem{deZP}
D. DeZeeuw and K. Powell, An Adaptively Refined Cartesian Mesh Solver for the Euler Equations,
{\em J. Computat. Phys.} {\bf 104}, 56 (1993).

\bibitem{schutz}B. Schutz, {\em A first course in general relativity} (Cambridge
University Press, New York, 1985).

\bibitem{martin} D. Martin and K. Cartwright, Solving Poisson's Equation using Adaptive Mesh Refinement, unpublished notes (1996).

\bibitem{teuk_lw} S. Teukolsky, Linearized quadrupole waves in general relativity and the motion of test particles Equation using Adaptive Mesh Refinement, {\em Phys. Rev.} {\bf D26}, 745 (1982).

\bibitem{OandP} F. Olsson and N. Anders Petersson, Stability of interpolation on 
overlapping grids, {\em Comput. Fluids} {\bf 25}, 583 (1996).

\end{thebibliography}
\end{document}